 \documentclass[12pt,preprint]{aastex}

\usepackage[lowtilde]{url}
\usepackage{amsmath}
\usepackage{color}
\usepackage{epstopdf}

\newcommand{\gsim}{ {}^>_\sim}
\newcommand{\lsim}{ {}^<_\sim}

\newcommand {\Lya}    {Ly$\alpha$}   
\newcommand {\Lyb}    {Ly$\beta$}    

\newcommand {\HI}     {\ion{H}{1}}   
\newcommand {\HII}    {\ion{H}{2}}   
\newcommand {\HeII}   {\ion{He}{2}}   
\newcommand {\HeIII}  {\ion{He}{3}}   

\newcommand {\etal}   {et~al.} 

\def\gsim{\lower 2pt \hbox{$\, \buildrel {\scriptstyle >}\over
{\scriptstyle \sim}\,$}}
\def\lsim{\lower 2pt \hbox{$\, \buildrel {\scriptstyle <}\over
{\scriptstyle \sim}\,$}}

\slugcomment{Accepted to ApJ (\today)}
\shorttitle{Critical SFR \& Reionization}
\shortauthors{Shull et al.}

\begin{document}

\title{Critical Star-Formation Rates for Reionization:  \\
   Full Reionization Occurs at Redshift $z \approx$ 7 }  

\author{J. Michael Shull, Anthony Harness,  Michele Trenti}
\affil{CASA, Department of Astrophysical \& Planetary Sciences, \\
University of Colorado, Boulder, CO 80309}
\author{Britton D. Smith, {\it Dept. of Physics \& Astronomy, Michigan State University} }
\email{michael.shull@colorado.edu, trenti@colorado.edu, anthony.harness@colorado.edu,
   smit1685@msu.edu}  


\begin{abstract} 
We assess the probable redshift ($z_{\rm rei} \approx 7$) for full reionization of the intergalactic medium 
(IGM) using a prescription for the co-moving star-formation-rate (SFR) density ($\dot{\rho}_{\rm SFR}$)
required to maintain photoionization against recombination.   Our newly developed on-line reionization 
simulator allows users to assess the required SFR and ionization histories, using a variety of assumptions
for galactic and stellar populations,  IGM clumping factor and temperature, and LyC escape fraction.
The decline in high-redshift galaxy candidates and \Lya\ emitters at $z = 6-8$ suggests a rising neutral 
fraction, with reionization at $z \gsim 7$ increasingly difficult owing to increased recombination rates and
constraints from the ionizing background and LyC mean free path.    The required rate is
$\dot{\rho}_{\rm SFR} \approx (0.018 \, M_{\odot}~{\rm yr}^{-1}~{\rm Mpc}^{-3})[(1+z)/8]^3 
(C_H/3)(0.2/f_{\rm esc}) T_4^{-0.845}$ scaled to fiducial values of clumping factor $C_H = 3$, escape 
fraction $f_{\rm esc} = 0.2$, electron temperature $T_e = 10^4$~K, and low-metallicity initial mass functions 
(IMF) and stellar atmospheres.  Our hydrodynamical + N-body simulations find a mean clumping factor 
$C_H \approx (2.9)[(1+z)/6]^{-1.1}$ in the photoionized, photoheated filaments at $z = 5-9$.   The critical 
SFR could be reduced by increasing the minimum stellar mass, invoking a top-heavy IMF, or systematically
increasing $f_{\rm esc}$ at high $z$.  
The CMB optical depth, $\tau_e = 0.088\pm0.015$, could be explained by full reionization,  producing 
$\tau_e = 0.050$ back to $z_{\rm rei}  \approx 7$, augmented by $\Delta \tau_e \approx 0.01-0.04$ in a 
partially ionized IGM at $z > 7$.  In this scenario, the strongest 21-cm signal should occur at 
redshifted frequencies $124-167$ MHz owing to IGM heating over an interval $\Delta z \approx 3$ 
from $z \approx 7.5-10.5$.  

\end{abstract} 


\section{INTRODUCTION}

Our current hypothesis is that the galaxies and black holes observed today originated over 13 Gyr 
ago, growing from seeds of primordial density perturbations. One can test this hypothesis by studying 
the star formation rate (SFR) history, from the Epoch of Reionization (EoR) at redshifts $6 < z < 12$, 
through the peak era at $z \approx 2-3$ (Hopkins \& Beacom 2006) down to the present epoch.   The 
feedback of ionizing radiation, kinetic energy, and heavy elements leaves imprints on early stars, 
supernovae, and galaxies, providing a ``fossil record" that can be detected through abundances in 
Galactic halo stars and the intergalactic medium (IGM) and in the distributions of mass, 
metallicity, and luminosity of galaxies.  

Determining when and how the universe was reionized by these early sources have been important 
questions for decades (Gunn \& Peterson 1965; Sunyaev 1977; Robertson \etal\ 2010).   It has been
suggested that IGM reionization was complete by $z \approx 6.5$ (Fan \etal\ 2001; Gnedin \& Fan 2006;  
Fan \etal\ 2006; Hu \& Cowie 2006), based on strong \Lya\ absorption from neutral hydrogen along lines 
of sight to QSOs at  $z > 6$.  Becker \etal\ (2007) and Songaila (2004) used transmission of the \Lya\
(and \Lyb) forest out to $z = 5.8$ and $z = 6.3$, respectively, to suggest a smoothly decreasing ionization 
rate toward higher redshifts.   Recent surveys of high-redshift galaxies and \Lya\ emitters (Bouwens \etal\ 
2011a;  Ouchi \etal\ 2010; Kashikawa \etal\ 2011; Ono \etal\ 2012; Schenker \etal\ 2012)  infer an increasing
IGM neutral fraction from the declining populations between $z = 6-8$.  Further evidence for an increasing
neutral fraction comes from the decreasing sizes of  ionized ``near zones" associated with quasars between 
$z = 5.7$ and $z = 6.4$ (Carilli \etal\ 2010) and from the \Lya\ damping wing in the transmission profile toward 
the newly discovered quasar at $z = 7.085$ (Mortlock \etal\ 2011).  These studies all suggest that the IGM is 
becoming increasingly neutral between $z = 6-7$, marking the end of cosmic reionization when ionized 
regions overlap and percolate.   Whether the epoch of full reionization occurs at $z_{\rm rei} \approx 7$ is
still not ascertained.  

A contrasting estimate of the EoR comes from the measured optical depth, $\tau_e  = 0.088\pm0.015$,
to electron scattering of the cosmic microwave background (CMB) in {\it WMAP-7} observations  
(Larson \etal\ 2011; Komatsu \etal\ 2011).  The error bars come from the central 68\% in the marginalized 
cumulative distribution.   Using additional cosmological parameter constraints, they infer
single-epoch reionization at $z_{\rm rei} = 10.6 \pm 1.2$.   Although such a high redshift could be explained 
with $\Lambda$CDM simulations and modeled SFR  histories (Choudhury \& Ferrara 2005; Trac \& Cen 2007; 
Shull \& Venkatesan 2008), the CMB observations are at variance with optical surveys that suggest late 
reionization, unless reionization is a process that extends to higher redshifts. 

For a fully ionized IGM, including both H and He, the optical depth $\tau_e = 0.050$ for $z_{\rm rei} = 7.0$ 
(see Section 2.1).  In the CMB analysis (Larson \etal\ 2011), marginalization of $\tau_e$ with other cosmological 
parameters allows the possibility of lower optical depth, with a reionization epoch as low as $z_{\rm rei} \sim 7-8$ 
at 95\% C.L.   One can also invoke a partially ionized IGM at $z_{\rm rei} > 7$, as discussed by many groups 
(Cen 2003; Venkatesan \etal\ 2003;  Ricotti \& Ostriker 2004;  Benson \etal\ 2006; Shull \& Venkatesan 2008).
However, even with the recent  progress in finding high-$z$ galaxies, we still do not know whether galaxies are the 
sole agents of reionization.  Current observations of high-$z$ galaxies leave open several ionization scenarios, 
some involving simple hydrogen reionization at $z \approx 10$ and \HeII\ reionization at $z \approx 3$, and 
others with more complex ionization histories (Bolton \& Haehnelt 2007;  Venkatesan, Tumlinson \& Shull 2003; 
Cen 2003) that depend on SFRs at $z = 7-20$.  Shull \& Venkatesan (2008) demonstrated how the CMB optical 
depth constrains the SFR and IGM metallicity history at $z > 7$, and Trenti \& Shull (2010) quantified the 
metallicity-driven transition from Population~III (metal-free stars) to Population~II (stars formed from 
metal-enriched gas).   

Several recent observations provide valuable constraints on the luminosity function of high-$z$ galaxies.  The 
number density of galaxies appears to drop rapidly at $z > 7$ (Bouwens \etal\ 2009, 2010a,b, 2011a,b,c).  With 
a comoving SFR density $\dot{\rho}_*  \lsim 0.01~M_{\odot}~{\rm yr}^{-1}~{\rm Mpc}^{-3}$ at $z \approx 7$
(Gonz\'alez \etal\ 2010; Bouwens \etal\ 2011a), the observed galaxies do not produce enough ionizing photons 
in the Lyman continuum (LyC) to maintain a photoionized IGM against recombinations.   However, the
luminosity function is steep, and the total LyC budget requires extrapolation to low-luminosity galaxies 
(Trenti \etal\ 2010;  Bouwens \etal\ 2011b).   Moreover, the conversion from SFR to LyC production rate
relies on insecure calibrations from theoretical models and comparison with high-mass, low-metallicity stars.   
We revisit the calculation of LyC photon production and assess the high-$z$ galaxy contribution 
to reionization.   We also analyze several factors, such as the photon escape fraction ($f_{\rm esc}$), 
IGM clumping factor ($C_H$), and electron temperature ($T_e$), which enter the calculation of the 
``critical star formation rate" ($\dot{\rho}_{\rm crit}$) necessary to maintain a photoionized IGM.  

In Section 2, we calculate $\dot{\rho}_{\rm crit}$ ($M_{\odot}~{\rm yr}^{-1}$ Mpc$^{-3}$) in a filamentary IGM,
equating the production rate of Lyman continuum (LyC) photons with the hydrogen recombination 
rate.   The photoionization rate depends on the mass function of stellar populations,  their evolutionary tracks 
and stellar atmospheres, and the escape fraction, $f_{\rm esc}$ of LyC photons away from their galactic 
sources.  The recombination rate depends on the density and temperature of the IGM, properties we explore 
with cosmological simulations.  In Section 3, we give our results for the critical SFR at $z \gsim 6$ and 
present our new SFR simulator, a user-friendly interface for calculating $\dot{\rho}_{\rm crit}(z)$ and $\tau_e(z)$. 
In Section 4, we discuss the implications for the hydrogen EoR.   Consistency between high-redshift galaxies and 
CMB optical depth appears to require $z_{\rm rei} \approx 7$ and a partially ionized IGM at $z > 7$.   
The peak signal from redshifted 21-cm emission would likely occur during the heating period between $z = 7.7-8.8$ 
(145--163 MHz) when the hydrogen neutral fraction $x_{\rm HI} \approx 0.5$ (Pritchard \etal\ 2010;  Lidz \etal\ 2008).

\section{REIONIZATION WITH CLUMPING AND PHOTON ESCAPE}

\subsection{Reionization and Critical Star Formation Rate}

We denote by $\dot{\rho}_{\rm SFR}$ ($M_{\odot}~{\rm yr}^{-1}~{\rm Mpc}^{-3}$) the global star formation 
rate per co-moving volume.   Using a simple argument (Madau \etal\ 1999), balancing photoionization with 
radiative recombination, we estimate the critical SFR to maintain IGM photoionization at $z > 7$, assuming
that the LyC photons are produced by populations of massive (OB-type) stars.
Because the mass in collapsed objects (clusters, groups, galaxies) is still small at high redshift,
the IGM contains most of the cosmological baryons, at mean density 
\begin{equation}
  \bar{ \rho}_b = \Omega_b \rho_{\rm cr} (1+z)^3 =  (4.24 \times 10^{-31}~{\rm g~cm}^{-3}) (1+z)^3   \; .   
\end{equation}
For a Hubble constant denoted $H_0 = (100~ {\rm km~s}^{-1}~{\rm Mpc}^{-1}) \, h$, we adopt the WMAP-7 
(plus BAO + $H_0$) parameters, $\Omega_b h^2 = 0.02255\pm0.00054$ and 
$\Omega_m h^2 = 0.1352\pm0.0036$ (Komatsu \etal\ 2011) relative to a critical density 
$\rho_{\rm cr} = 1.8785 \times 10^{-29}~h^2$ g~cm$^{-3}$.  
From the corresponding helium mass fraction $Y = 0.2477\pm0.0029$ (Peimbert \etal\ 2007), we adopt 
a mean hydrogen number density,
\begin{equation} 
     \bar{n}_H = \frac { \bar{\rho}_b (1-Y)}{m_H} = (1.905 \times 10^{-7}~{\rm cm}^{-3}) (1+z)^3      \; .  
\end{equation} 

In a fully ionized IGM, the CMB optical depth back to $z_{\rm rei}$ can be written as the 
integral of $n_e \sigma_T d \ell$, the electron density times the Thomson cross section along proper length,   
\begin{equation}
   \tau_e(z_{\rm rei}) = \int _{0}^{z_{\rm rei}} n_e \sigma_T (1+z)^{-1} \; [c/H(z)] \; dz    \; ,
\end{equation}
for a standard $\Lambda$CDM cosmology ($\Omega_m + \Omega_{\Lambda} = 1$) with 
$H(z) = H_0 [\Omega_m (1+z)^3 + \Omega_{\Lambda}]^{1/2}$.
This integral can be done analytically (Shull \& Venkatesan 2008),
\begin{equation}
   \tau_e(z_{\rm rei}) = \left( \frac {c \, \sigma_T}{H_0} \right) \left( \frac {2 \Omega_b}  {3 \Omega_m} \right)
     \left[ \frac {\rho_{\rm cr} (1-Y)(1+y) } {m_H} \right]
     \left[ \{ \Omega_m (1+z_{\rm rei})^3 + \Omega_{\Lambda} \}^{1/2} - 1 \right] \; .  
\end{equation}
In the high-redshift limit, when $\Omega_m (1+z)^3 \gg \Omega_{\Lambda}$, this expression simplifies to 
\begin{equation}
      \tau_e(z_{\rm rei})  \approx \left( \frac {c \, \sigma_T}{H_0} \right) \left( \frac {2 \Omega_b}
      {3 \Omega_m^{1/2}} \right)  \left[ \frac {\rho_{\rm cr} (1-Y)(1+y)}
      {m_H} \right] (1+z_{\rm rei})^{3/2} \approx (0.0521) \left[ \frac {(1+z_{\rm rei})}{8} \right] ^{3/2}  
\end{equation}    
independent of $h$ to lowest order.  The helium and electron densities are written $n_{\rm He} = y n_H$ and 
$n_e = n_H (1+y)$ for singly ionized helium, where $y = n_{\rm He}/n_{\rm H} = (Y/4)/(1-Y) \approx 0.0823$ 
by number.   To these formulae, we add  $\Delta \tau_e \approx 0.002$, from electrons donated 
by \HeIII\ reionized at $z \leq 3$ (Shull \etal\ 2010).  Helium therefore contributes $\sim$8\% to $\tau_e$,
and a fully ionized IGM produces $\tau_e = 0.050$, 0.060, and 0.070 back to redshifts $z_{\rm rei} = 7$, 8, 
and 9, respectively.   

A comoving volume of 1 Mpc$^3$ contains $N_H = 5.6 \times 10^{66}$ hydrogen atoms.  Our simple ionization 
criterion requires a SFR density that produces a number of LyC photons equal to $N_H$ over a
hydrogen recombination time, $t_{\rm rec} = [n_e \alpha_H^{(B)} C_H]^{-1}$.  The hydrogen Case-B 
recombination rate coefficient (Osterbrock \& Ferland 2006) is
$\alpha_H^{(B)}(T) \approx (2.59 \times 10^{-13}~{\rm cm}^3~{\rm s}^{-1}) T_4^{-0.845}$, scaled to an IGM 
temperature $T = (10^4~{\rm K})T_4$.   
For typical IGM ionization histories and photoelectric heating rates, numerical simulations predict that
diffuse photoionized filaments of hydrogen have temperatures ranging from 5000~K to 20,000~K 
(Dav\'e \etal\ 2001; Smith \etal\ 2011).  These are consistent with temperatures inferred from observations 
of the \Lya\ forest at $z < 5$ (Becker \etal\ 2011).

Owing to gravitational instability, a realistic IGM is inhomogenous and filamentary.  Semi-analytical 
models of the reionization of the universe often adopt a ``clumping factor", 
$C_H \equiv  \langle n_e^2 \rangle / \langle n_e \rangle^2$,  to account for 
inhomogeneity in estimates of the enhanced recombination rate in denser IGM filaments.  The clumping factor 
therefore plays an important role in computing the critical SFR density needed to maintain the reionization of the 
universe.   The clumping factor is also used in numerical simulations to implement ``sub-grid physics'', in which
changes in the density field occur on scales below the resolution of the simulation and are also approximated by
 the factor $C_H$ (Gnedin \& Ostriker 1997; Madau \etal\ 1999;  Miralda-Escud\'e \etal\ 2000; Miralda-Escud\'e 2003;  
Kohler \etal\ 2007).   

We correct the recombination time for density variations scaled to a fiducial $C_H \approx 3$,  found in the 
simulations described below.   At $z \approx 7$, the IGM filaments have electron density 
$n_e \approx (10^{-4}~{\rm cm}^{-3})[(1+z)/8]^3 C_H$, and the characteristic times for hydrogen 
recombination and Hubble expansion are,
\begin{eqnarray} 
  t_{\rm rec} &\approx& (386~{\rm Myr})(3/C_H) T_4^{0.845} \left[ \frac {(1+z)}{8} \right]^{-3}  \;, \\
   t_{\rm H}   &\approx& [H_0 \Omega_m^{1/2} (1+z)^{3/2}]^{-1} \approx (1.18~{\rm Gyr})
                   \left[ \frac {(1+z)}{8} \right]^{-3/2}  \;  .   
\end{eqnarray} 
In our calculations, we express the reionization criterion as 
$N_H ({\rm Mpc}^{-3}) = \dot{\rho}_{\rm crit} t_{\rm rec} Q_{\rm LyC} f_{\rm esc}$, where $\dot{\rho}_{\rm crit}$ is 
the critical SFR density ($M_{\odot}~{\rm yr}^{-1}~{\rm Mpc}^{-3}$) and $Q_{\rm LyC}$ is the conversion factor 
from $\dot{\rho}_{\rm SFR}$ to the LyC production rate (see Section 2.2).  
We define $f_{\rm esc}$ as the fraction of LyC photons that escape from their galactic 
sources into the IGM (Dove \& Shull 1994).   Recent statistical estimates (Nestor \etal\ 2011) suggest that 
$f_{\rm esc} \approx 0.1$ for an ensemble of 26 Lyman-break galaxies and 130 Ly$\alpha$ emitters at 
$z \approx 3.09\pm0.03$, and it could be higher for the lower-mass galaxies that likely dominate the 
escaping LyC at $z > 6$ (Fernandez \& Shull 2011).
The LyC production efficiency, $Q_{\rm LyC}$, is expressed in units $10^{63}$ photons per $M_{\odot}$ 
of star formation, since typical massive stars emit $(1-10) \times 10^{63}$ LyC photons over their lifetime.
To evaluate $Q_{\rm LyC}$, we convert the SFR (by mass) into numbers of OB-stars and compute 
the total number of ionizing photons produced by a star of mass $m$ over its lifetime.  We then
integrate over an IMF, $\Psi(m) = K m^{-\alpha}$, with a range $m_{\rm min} < m < m_{\rm max}$.
The standard mass range is 0.1 $M_{\odot}$ to 100 $M_{\odot}$, but  changes in the mass range and IMF 
slope will affect the LyC production substantially.  For example, differences in the IMF have been associated 
with higher Jeans masses in low-metallicity gas in the high-redshift IGM (Abel \etal\  2002; Bromm \& Loeb 2003),
and Tumlinson (2007) and Smith \etal\ (2009) noted the potential 
influence of CMB temperature on modes of high-redshift star formation.  Consequently, most cosmological 
simulations or calculations include a metallicity-induced IMF transition (Trenti  \& Shull 2010) between 
low-metallicity Population~III star formation and metal-enhanced Population~II.

\subsection{Ionizing photon luminosities} 

The number of LyC photons produced per total mass in star formation depends on the IMF of the stellar 
population and is given by the conversion factor $Q_{\rm LyC}$.    We calculate this conversion by integrating 
the total number of LyC photons produced over the entire mass in star formation. 
\begin{equation}
    Q_{\rm LyC} \equiv \frac{N_{\rm LyC}}{\dot{\rho}_{\rm SFR}t_{\rm rec}} = \frac{\int_{m_{OB}}^{m_{max}} 
        \Psi (m) Q(m)  \,dm} {\int_{m_{min}}^{m_{max}} \Psi (m)  m \, dm}    \; . 
\end{equation}
Here, $Q(m)$ is the lifetime-integrated number of  LyC photons as a function of mass calculated from stellar 
atmosphere models and evolutionary tracks.   The IMF, $\Psi (m) = Km^{-\alpha}$, is  integrated over the mass range 
$m_{\rm min} < m < m_{\rm max}$, where $m$ is expressed in solar units.  In the SFR simulator discussed in
Section 3.3, the user can choose between a normal and broken IMF power law. The lower integration limit in 
the numerator, $m_{\rm OB}$, is the mass at which stars no longer produce significant amounts of LyC photons. 

In our calculator, we use and compare two models that calculate $Q(m)$.  First, using stellar atmospheres and 
evolutionary tracks (R.\ S.\ Sutherland \& J.\ M.\ Shull, unpublished), we find that, over its main-sequence and 
post-main-sequence lifetime, an OB star of mass $m$ produces a total number $N_{\rm LyC} = Q_{63}(m) \times 10^{63}$ 
of ionizing photons, where $Q_{63} \approx 1-10$ over the mass range $m = 30-100$ and for metallicities 
$Z  =( 0.02 - 2.0) Z_{\odot}$.  We have fitted our results to the form $Q_{63}(m) \approx  Am - B$, for 
$m \geq m_{\rm OB} = B/A$ (the mass $m_{\rm OB}$ defines the effective lower limit for stars that produce significant 
numbers of LyC photons).  For metallicities in the range $0.002 \leq Z \leq 0.04$ (where $Z = Z_{\odot} = 0.02$ is 
the solar metallicity) and for masses $15 < m < 100$,  the fitted coefficients are $B = 1.578$ and $A(Z) = 0.0950 - 1.059Z$.  
Thus, for $Z = Z_{\odot} \approx 0.02$, we have $A = 0.0738$ and $m_{\rm OB} = 21.4\,M_{\odot}$, while for 
$Z = 0.1 Z_{\odot} = 0.002$ we have $A = 0.0929$ and $m_{\rm OB} = 17\,M_{\odot}$.   Inserting these values for $Q(m)$ 
into equation (8), we integrate these LyC photon yields over the IMF, to derive the conversion coefficient, $Q_{\rm LyC}$, 
from {\it total mass} in star formation to {\it total number} of LyC photons produced (in units of $10^{63}$ 
photons per $M_{\odot}$ ).  

The integrals in eq.\ (8) can be done analytically as functions of the IMF parameters and metallicity. 
For a Salpeter IMF ($\alpha = 2.35$, $0.1 \leq m \leq 100$) we find: $Q_{\rm LyC} = 0.00236$ 
(for $Z = Z_{\odot}$),  $Q_{\rm LyC} = 0.00366$ (for $Z = 0.2 Z_{\odot}$),  and $Q_{\rm LyC} = 0.00384$ 
(for $Z = 0.1 Z_{\odot}$).  Adopting a typical (low-$Z$) value $Q_{\rm LyC} = 0.004$ ($4 \times 10^{60}$ 
photons per $M_{\odot}$), we can solve for the critical SFR for reionization, scaled to clumping factor
$C_H = 3$, escape fraction $f_{\rm esc} = 0.2$, and gas temperature $T_4 = 1$:    
\begin{equation}
  \dot{\rho}_{\rm crit} = (0.018~M_{\odot}~{\rm yr}^{-1}~{\rm Mpc}^{-3})
   \left[ \frac {(1+z)}{8} \right]^{3} 
   \left[ \frac {C_H/3} {f_{\rm esc}/0.2} \right]
   \left[ \frac {0.004} {Q_{\rm LyC}} \right] T_4^{-0.845} \; .  
\end{equation}
Our chosen value of $C_H \approx 3$ is consistent with recent downward revisions (Pawlik \etal\ 2009) 
and with our numerical simulations discussed in Section 2.3.  Escape fractions $f_{\rm esc} \approx 0.1-0.2$
have been inferred from observations at $z \approx 3$ (Shapley \etal\ 2006; Nestor \etal\ 2011) and
theoretical expectations (Fernandez \& Shull 2011).  
Equation (9) agrees with prior estimates (Madau \etal\ 1999) when adjusted for our new scaling factors,
in particular the ratio $(C_H / f_{\rm esc} = 15$).  Their earlier paper assumed $C_H = 30$, 
$f_{\rm esc} = 1$, $\Omega_b h^2 = 0.020$, $Q_{\rm LyC} = 0.005$, and $z = 5$.   
Our fiducial redshift has increased from $z = 5$ to $z = 7$, appropriate for the new discoveries of 
high-redshift galaxy candidates for reionization.  Expressed with the same coefficients in eq.\ (9), their 
coefficient would be nearly the same,  $0.020\,M_{\odot}~{\rm yr}^{-1}~{\rm Mpc}^{-3}$ at $z = 7$.   
One of the improvements in our formulation is to better identify the dependences on the physical 
parameters ($C_H$, $f_{\rm esc}$, $T_e$) and the SFR-to-LyC conversion factor ($Q_{\rm LyC})$,
which can change with different IMFs and atmospheres.  

A related calculation is the production rate of ionizing (LyC) photons per unit volume, needed to balance 
hydrogen recombinations.  With the same assumptions as above, this is
\begin{equation}
  \left( \frac {d {\cal N}_{\rm LyC}} {dt} \right)_{\rm crit} = 
       n_e \, n_{\rm HII} \, \alpha_H^{(B)} =  (4.6 \times 10^{50}~{\rm s}^{-1}~{\rm Mpc}^{-3}) 
       \left[ \frac {(1+z)}{8} \right]^3 T_{4}^{-0.845}    \left( \frac {C_H}{3} \right)   \;  .
\end{equation} 
For standard high-mass stars (O7~V, solar metallicity) each with LyC photon luminosity $10^{49}$~s$^{-1}$, 
this rate corresponds to an O-star space density $n_O  \approx 50~{\rm Mpc}^{-3}$ at $z = 7$.  
Schaerer (2002, 2003) also computed models of Population III and low-metallicity stars based on 
non-LTE model atmospheres and new stellar evolution tracks and evolutionary synthesis models. 
The lifetime total number of LyC photons produced per star of mass $m$ is 
$Q(m) = \bar{Q}(H) \times t_\ast$. For stars of mass parameter $x = \log (m/M_\odot)$, the 
number of ionizing photons, $\bar Q$, emitted per second per star, is given by:
\begin{equation}
\text{log}_{10}\left[\bar{Q}_H/s^{-1}\right]	= \left\{
\begin{array}{lll}
	43.61 + 4.90x - 0.83x^2	&	Z = 0,	&	9-500 M_\odot, \\
	39.29 + 8.55x	&	Z=0,	&	5-9 M_\odot,\\
	27.80+30.68x-14.80x^2+2.50x^3	& Z = 0.02 Z_\odot, &	7-150 M_\odot, \\
	27.89+27.75x-11.87x^2+1.73x^3	&	Z= Z_\odot,	&	7-120 M_\odot \\
\end{array}		
\right.
\label{eq:Qvalues}
\end{equation}

\noindent 
and the star's lifetime $t_\ast$ is given by:

\begin{equation}
\text{log}_{10}\left[t_\ast\right]	= \left\{
\begin{array}{ll}
	9.785-3.759x+1.413x^2-0.186x^3	&	Z = 0, \\
	9.59-2.79x+0.63x^2	&	Z = 0.02 Z_\odot, \\
	9.986-3.497x+0.894x^2	&	Z = Z_\odot
\end{array}		
\right.
\label{eq:lifetimes}
\end{equation}

Sample values of $Q_{\rm LyC}$ and $\dot{\rho}_{\rm crit}$ for different IMFs are shown in Table~1,
for both model atmospheres.    One can obtain different values of $Q_{\rm LyC}$ for power-law IMFs
by varying their high-mass slope $\beta$ and the minimum and maximum masses.  Increases in 
$Q_{\rm LyC}$ translate into  {\it decreases} in critical SFR.  For example, at $0.1\,Z_{\odot}$, if the 
minimum mass is raised to $1\,M_{\odot}$ with a Salpeter slope ($\beta = 2.35$), 
the LyC production factor rises to $Q_{\rm LyC} = 0.0098$, some 2.5 times higher than the equivalent 
model, $Q_{\rm LyC} = 0.00384$ for $m_{\rm min} = 0.1 M_{\odot}$.  If the IMF is flatter, $\beta = 2$ 
instead of 2.35,  with minimum mass fixed at $m_{\rm min} = 0.1 M_{\odot}$, $Q_{\rm LyC} = 0.0177$ 
(4.6 times higher). At the upper end of the IMF, if one increases $m_{\rm max}$ from 100 to 200 $M_{\odot}$, 
keeping $\beta = 2.35$ and $m_{\rm min} = 0.1$, one finds $Q_{\rm LyC} = 0.0054$ (1.4 times higher). 
All of these variations can be explored with our reionization/SFR simulator, described in Section 3.3.

The critical SFR of  $0.018\,M_{\odot}~{\rm yr}^{-1}~{\rm Mpc}^{-3}$ can be understood from simple 
arithmetic.   Over the 386 Myr recombination time (eq.\ [6]) with $C_H = 3$,  $f_{\rm esc} = 0.2$, and $z = 7$, 
approximately 20 million stars are formed per comoving Mpc$^3$ in a Salpeter IMF ($0.1-100\,M_{\odot}$)
with mean stellar mass $0.352 M_{\odot}$.  Of these stars, a small fraction, $f_{\rm OB} \approx 7 \times 10^{-4}$, 
are massive OB stars ($m > 20 M_{\odot})$.  At an average of $2\times10^{63}$ LyC photons and escape fraction 
$f_{\rm esc} = 0.2$, these $\sim$14,000 OB stars will produce a net (escaping) $\sim 5\times10^{66}$ LyC 
photons over their lifetime.   These photons are sufficient to ionize $N_H = 5.6 \times 10^{66}$ hydrogen 
atoms Mpc$^{-3}$.  

\newpage
 
\subsection{Clumping Factor from  Cosmological Simulations}

The simulations used in this study were performed with \texttt{Enzo}, an Eulerian adaptive mesh-refinement (AMR), 
hydrodynamical + N-body code (Bryan \& Norman 1997;  O'Shea \etal\ 2004, 2005).  Smith \etal\ (2011) enhanced 
this code by adding new modules for star formation, primordial chemistry, and cooling rates consistent with ionizing 
radiation, metal transport, and feedback. The ionizing background is spatially constant and optically thin, but 
variable in redshift.  At this stage, we have not implemented radiative transfer or spectral filtering by the IGM.  

For the clumping factor calculation, we ran a simulation on a 50$h^{-1}$~Mpc static grid (``unigrid") cube with 
$1024^3$ cells, denoted as run 50\_1024\_2 in Table~1 of  Smith \etal\ (2011).   
The radiative heating from the ionizing background plays an important role in determining the properties of the 
filamentary structure.   As a filament is ionized and heated, its density drops and its temperature rises;  both
effects reduce the recombination rate.   To study these effects, we ran four moderate-resolution (50$h^{-1}$~Mpc 
unigrid cube with $512^3$ cells) simulations with varying ionizing backgrounds, summarized in Table~2. 
After initial submission of this paper, we ran a $1536^3$ simulation, to check convergence and assess the
``cosmic variance" among eight sub-volumes of the $1024^3$ and $1536^3$ simulations.   
The standard UV background was taken from Haardt \& Madau (2001), although we also explore new computations 
of high-$z$ SFRs by Trenti \etal\ (2010) and Haardt \& Madau (2012).  
These four simulations were:  (1)  no photoionizing background;  (2)  UV background ramped up from $z=7$ to $z=6$ 
(run 50\_512\_2 from Smith \etal\ 2011);  (3) UV background ramped up from $z=9$ to $z=8$; and
(4) UV background ramped up from $z=9$ to $z=8$, but twice as strong as in (3).  
Post-processing of the simulations was performed using the data analysis and visualization package, 
\texttt{yt}\footnotemark, documented by Turk \etal\ (2011).\footnotetext{http://yt.enzotools.org/}

For regions of ionized hydrogen of density $n_{\rm HII}$, we calculate the clumping factor, $C_H$.
using two different methods.   The first calculation, which has been used in some earlier studies, uses
density weighting.   In this ``density field" (DF) method we define
\begin{equation}
   C_{\rm DF} = \frac { \langle n_{\rm HII}^2 \rangle } { \langle n_{\rm HII} \rangle^2}    \; , 
\end{equation} 
and average the density fields (quantities $x_i$, denoting either $n_{\rm HII}$ or $n_{\rm HII}^2$) where
parameter averages are computed by summing over grid cells ($j$), subject to various ``cuts'' on the IGM overdensity 
and gas temperature and weighted by factors, $w_j$, 
\begin{equation}
	\left<x_i\right> =\displaystyle\sum\limits_j x_j w_j / \displaystyle\sum\limits_j w_j   \; .
	\label{eq:average}
\end{equation}
In the second calculation, we compare the local recombination rate to the global average recombination rate, averaged
over density and temperature in cells, 
\begin{equation}
	C_{\rm RR} = \frac {\langle  n_e \, n_{\rm HII} \;  \alpha_H^{(B)} (T)  \rangle }   
	          { \langle  n_e \rangle \langle  n_{\rm HII} \rangle  \, \langle  \alpha_H^{(B)} (T) \rangle}  \; , 
\end{equation}
where again $\langle x \rangle$ denotes a weighted average and $\alpha_H^{(B)} (T)$ is the case-B radiative 
recombination rate coefficient for hydrogen, as tabulated by Osterbrock  \& Ferland (2006).  We refer to this as the 
``recombination rate" (RR)  method.  

Because the purpose of the clumping factor is to correct for an enhanced recombination rate, we believe $C_{\rm RR}$
to be a more appropriate representation.  The recombinations that are important to removing ionizing photons occur in 
the filamentary structure of the IGM, and the clumping factor should only be calculated in these structures. 
To assess the critical SFR necessary to maintain an ionized medium, we focus on grid cells that are significantly 
ionized.  Because a negligible amount of recombination occurs in cells containing mostly neutral gas, these cells should 
not contribute to the clumping factor. If we do not exclude neutral gas, the clumping factor is extremely high ($C_H \sim 100$) 
before the ionizing background is turned on.  We find large density gradients between the neutral gas (uniformly distributed 
over the simulation) and the ionized gas.  
Because the clumping factor is a measure of the inhomogeneity of the medium, a large density gradient yields a large
value of $C_H$.  If low-density voids are included in the calculation, the larger density gradient leads to an overestimate 
of the clumping.   By setting both upper and lower density thresholds, we can exclude collapsed halos and low-density 
voids from our calculations.  

Previous studies (Miralda-Escud\'e 2000; Miralda-Escud\'e \etal\ 2003; Pawlik \etal\ 2009) addressed these issues by 
setting a density threshold that excludes collapsed halos, but they did not set a lower limit to exclude voids from their 
calculations. To explore these effects in a filamentary IGM, we make various cuts of our data in baryon overdensity
($\Delta_b \equiv \rho_b/\bar{\rho}_b$) and in temperature, metallicity, and hydrogen ionization fraction 
($x \equiv n_{\rm HII}/n_H$).   In our standard formulation, we include only those cells that meet the following criteria:  
$1 < \Delta_b < 100$, $300~{\rm K} < T < 10^5$~K, $Z < 10^{-6}\, Z_\odot$,  and $x > 0.05$.  We believe this ``data cut" 
adequately represents unenriched IGM lying on an adiabat  (see Figure~19 of Smith \etal\ 2011).  We also explore 
the clumping factor with no lower density threshold (total range $\Delta_b < 100$).     Our results, presented in Section 3.1, 
show a small increase in $C_H(z)$ from the wider range in densities by including low-density cells with $\Delta_b < 1$.  
However, these low-density voids do not contribute substantially to the recombination rate.

In summary, our prescriptions for calculating the clumping factor yield a more physical representation of the enhanced 
recombination rate that is an important component to many reionization models. By not assuming a fully ionized medium 
and by specifically following $n_{\rm HII}$, we are able to exclude denser neutral gas that does not contribute appreciably 
to the recombination rate.  We make simple assumptions on the reionization process and reionization history (Section 3.2), 
turning on the ionizing radiation field at redshifts $z = 7$ or $z = 9$ and following the thermal history arising from 
photoelectric heating, radiative cooling, and pressure smoothing (Pawlik \etal\ 2009).  The metal-line and molecular cooling, 
metal transport, and feedback included in our simulations also allow us to accurately represent the thermodynamics of the 
gas, which has been shown to have a significant effect on the evolution of the clumping factor.   Future models will include 
discrete sources and radiative transfer, accounting for temperature increases arising from photo-heating with spectral 
hardening (Abel \& Haehnelt 1999).  Because our current simulations employ a spatially constant ionizing radiation field, 
we anticipate carrying out these more realistic situations.


\section{RESULTS FROM SIMULATIONS}  

\subsection{Computing the Clumping Factor}

Early studies of IGM clumping adopted high values, $ C_H > 40$ at $z < 5$ (Gnedin \& Ostriker 1997). 
As discussed earlier, we believe these values are too high for the ionized IGM filaments, which have expanded 
as a result of the heat deposited by LyC photons.  The differences between observations and inferred critical 
SFR densities can largely be attributed to this high clumping factor (Sawicki \& Thompson 2006;
 Bouwens \etal\ 2007).  More recent studies have trended towards less clumping.   
 Rai{\v c}evi{\'c}  \& Theuns (2011) argue that using a global clumping factor overestimates
the recombination rate, and that local values should be used instead. In this study, we calculate the clumping 
factor for a series of high-resolution cosmological simulations for ionized hydrogen and helium that explore how the 
photoheating of an ionizing background affects the IGM thermodynamics (density and temperature). 
We impose criteria in overdensity, temperature, metallicity, and ionization fraction to constrain our calculations 
to IGM filaments where recombinations and ionization fronts are most important. 
The clumping in simulation $50\_1024\_7$ is calculated as a function of redshift from $z=15$ to $z=0$ for both 
$C_{\rm DF}$ and $C_{\rm RR}$, weighted by overdensity and by volume.  Note that the unigrid simulations
automatically yield volume weighting, when averaged over cells.   
For the remainder of our simulations, we find a power-law overdensity distribution, with an average form 
$f(\Delta_b) = 10^6 \, \Delta_b^{-1.5}$.  Weighting these cells by overdensity gives undue emphasis on small numbers 
of high-density cells.  This method is also biased, since it does not calculate the clumping factor where most of the 
recombinations are occurring. 

Figure 1 compares the DF and RR methods and shows the difference between the different weights used 
in our calculations. In both methods, weighting by overdensity results in higher clumping factors.   
The recombination-rate method (eq.\ [15]) results in a lower clumping factor because it accounts for both density and
temperature effects from the ionizing background on the  clumping averages.  Photoelectric heating during photoionization 
causes the filaments to expand, lowering the density.  The increased electron temperature also reduces the radiative 
recombination rate coefficient.   The DF method (eq.\ [13]) includes no dependence on the recombination rate coefficient, 
and therefore only counts the effect of photoheating on the filament density.  See Section~3.2 for further discussion.

In our current simulations, we turn on a spatially constant UV background at $z = 7$ or $z = 9$.  The effects of different 
``density cuts" in the summation over cells (see Section 2.3) are shown in  Figure 2.   We find little difference at $z > 9$ 
(turn-on of ionizing radiation), and see a small increase in $C_H$ at $z < 9$ when we include the lower density cells 
with $\Delta_b < 1$.  In future, more refined simulations with radiative transfer, we expect photoionization to be initiated 
in high-density regions, where the stars are located.   Once the UV background is turned on, lower-density regions are 
easily photoionized.    
The photoelectric heating of the UV background causes the filaments to expand and become diffuse, resulting
in a lower clumping factor.  The elevated temperature also reduces the recombination rate coefficient,
which in turn lowers $C_H$ when calculated by the RR method.  When calculating a mean recombination 
rate, only the recombining ionized gas is relevant.   By $z\sim6$, the IGM 
is nearly completely ionized and only small regions of  \HI\ remain.  As noted earlier, in more refined simulations,
the remaining \HI\ would be largely self-shielded from the UV background and likely to reside in clumps of 
high density yet unreached by ionization fronts.    

Figure 3 explores the convergence and cosmic variance among different simulations.   In two panels, we compare 
results from our $512^3$ simulations with larger simulations with $1024^3$ and $1536^3$ cells.  In each panel, we 
show the average clumping factors, together with those in eight sub-volumes.  The average values of $C_H$ agree 
in the $1024^3$ and $1536^3$ simulations, and the small ($\pm10$\%) variance 
among these sub-volumes suggests that the $1024^3$ and $1536^3$ simulations are converged.  Over the redshift 
range $5 < z < 9$ in the $1536^3$ simulation, the clumping factor is well fitted by a power-law,
 \begin{equation}
    C_H(z) = (2.9) \left[ \frac {(1+z)}{6}  \right] ^{-1.1}   \;  ,
\end{equation}
representing a slow rise in clumping to lower redshift, after the turn-on of ionizing radiation.

\subsection{Ionizing Background Study}

To study the effect of radiative heating from the ionizing background on filamentary structure, we ran a suite of 
moderate-resolution simulations with different ionizing backgrounds (see Table~2 for details).   The redshift
at which the UV background is turned on affects the clumping factor, causing it to drop and then recover at lower
redshift. 
Pawlik \etal\ (2009) attribute this effect to \emph{Jeans filtering}, where the photo-heating raises the cosmological 
Jeans mass, preventing further accretion onto low-mass halos and smoothing out small-scale density fluctuations. 
The photoheating also heats the filaments we are focusing on, which lowers the hydrogen recombination rate
coefficient, $\alpha_H^{(B)} \propto T^{-0.845}$, and causes the filaments to expand;  both of these effects reduce the 
clumping factor. Without a UV background, the clumping continues to rise as the filaments gravitationally collapse.
Pawlik \etal\ (2009) also claim that the clumping factor at $z=6$ is insensitive to when the background 
is turned on, as long as it is turned on at $z>9$.    When an ionizing background is introduced, photoheating acts 
as a positive feedback to reionization by lowering the clumping factor and making it  easier to stay ionized. This 
same photoheating mechanism also suppresses star formation in low-mass halos, which in turn lowers the ionizing 
photon production rate by star-forming galaxies and acts as a negative feedback to reionization (Pawlik \etal\ 2009).   
These thermodynamic processes affect clumping and structure and emphasize the importance of carefully 
modeling the strength of feedback processes and their effects on Jeans mass.

We have explored what happens when the ionizing radiation turned on earlier, especially during the interval 
$z = 6-8$ marking the transition from a neutral to fully ionized IGM.  In a series of $512^3$ simulations, 
we explore the influence of turn-on of photoionizing radiation between redshifts $z = 7$ and $z = 9$.  
Figure~4 shows $C_H$, computed for the moderate-resolution simulations via the density-field method and 
weighted by volume. We do not plot simulation $50\_512\_9\_2$, since it is identical to simulation $50\_512\_9$. 
As in the results of Pawlik \etal\ (2009), we find that photoheating from the ionizing background results in a 
decrease in the clumping factor.  The clumping factors fall along two tracks:  a higher track at redshifts above the 
turn-on of the UV background, and a lower one after turn-on.  
By $z = 5$, we find nearly identical clumping factors for both backgrounds.  After a substantial recovery time, the 
redshift when the background is turned on is not important.  Before this recovery, the clumping factor of the earlier 
background is lower by a factor of $\sim$2, resulting in earlier reionization. 

One can compare the strengths of feedback to reionization, where ``positive feedback" lowers the clumping factor 
and ``negative feedback"  suppresses star formation.  
At $z=5$ the total stellar mass (SFR density) of simulation $50\_512\_7$ is 1.19 (1.13) times lower than that of 
simulation $50\_512\_0$, while the clumping factor is 1.64 times lower. For simulation $50\_512\_9$, the total stellar 
mass (SFR density) and clumping factor are 1.52 (1.49) and 1.66, respectively, times lower than those of simulation 
$50\_512\_0$. This suggests that the positive feedback introduced by the background is greater than the negative 
feedback. However, the stellar mass (SFR density) does not recover in the same manner as the clumping factor.
Once photoheating suppresses the formation of small-mass halos, the Hubble flow takes over and prevents them from 
collapsing and forming stars. Therefore, the redshift at which the background is turned on determines whether the 
positive or negative feedback dominates and whether the ionizing background will cause reionization to be 
accelerated or delayed.

\subsection{Reionization SFR Simulator} 

In connection with this project, we have developed a user interface for calculating the critical SFR density 
$\left(\dot{\rho}_{\rm crit}\right)$ needed to maintain the IGM ionization at a given redshift. The software computes 
the effects of variations in the stellar IMF (slope, mass-range) and model atmospheres, and the redshift evolution of 
metallicity and gas thermodynamics (density, temperature, coupling to the CMB). The clumping factor and LyC escape 
fraction are free parameters in the calculator.  Our simulations find ranges of $C_H \approx 1 - 10$ depending 
on redshift, overdensity, and thermal phase of the (photoionized or shock-heated) IGM.   For the new
$1536^3$ simuation, the global mean clumping factor is $\langle C_H \rangle \approx 3$, with a power-law 
fit  $C_H(z) = (2.9) [(1+z)/6]^{-1.1}$ for redshifts between $5 < z < 9$.  

This calculator is a useful tool for determining the population of galaxies responsible for reionization. The critical 
SFR per co-moving volume (Eq. [9]) is obtained by balancing the production rate of LyC photons 
with the number of hydrogen recombinations.  Here, $C_{\rm H}$ is the clumping factor of ionized 
hydrogen, $f_{\rm esc}$ is the escape fraction of LyC photons from their host galaxies, and $T_4$ is the temperature 
scaled to $10^4$~K.   The conversion factor, $Q_{\rm LyC}$, from stellar mass to total number of LyC
photons produced, is regulated by the IMF and model atmospheres.   The user has the option of controlling these 
parameters to determine the resulting critical SFR density, subject to several observational constraints.
This simulator can be accessed at  \url{http://casa.colorado.edu/~harnessa/SFRcalculator} with an \texttt{html} 
interface for easy use.  

Two standard tests of the SFR use the simulator to calculate the ionization histories of \HII\ and \HeIII\ and compare
them to the ionized volume filling factor, $Q_{\rm HII}(z)$, and the CMB optical depth, $\tau_e(z)$.
The calculator derives $\tau_e(z)$ by integrating the differential form of Eq.\ (3) for various SFR histories and 
parameters.  The average evolution of $Q_{\rm HII}$ is found by numerical integration of the rate equation (Madau, 
Haardt, \& Rees 1999) expressing the sources and sinks of ionized zones,
\begin{equation}
	\frac{dQ_{\rm HII}}{dt} = \frac {\dot{n}_{\rm LyC}} {\langle n_H \rangle} - 
	       \frac{Q_{\rm HII}}{\langle t_{\rm rec}(C_H) \rangle}  \;  .
\label{eq:ionization_history}
\end{equation}
The source term, $\dot{n}_{\rm LyC} = \dot{\rho}_{\rm SFR}  f_{\rm esc} Q_{\rm LyC}$, represents the net ionizing photon 
production rate, computed from selected models of the SFR density, $\dot{\rho}_{\rm SFR}$.  Here, 
 $\langle n_H \rangle$ is the mean hydrogen number density, and $\langle t_{\rm rec} \rangle$ is the 
hydrogen recombination timescale, which depends on $C_H$ as shown in Eq.~(6). The IGM is assumed to be fully 
ionized when $Q_{\rm HII} = 1$.   We integrate a similar equation for $Q_{\rm HeIII}$ to follow  \HeII\ photoionization in the 
QSO 4-ryd continuum.  We use the QSO emissivities at 1 ryd  (Haardt \& Madau 2012), extrapolated to 4 ryd assuming 
a spectrum with specific flux $F_{\nu} \propto \nu^{-1.8}$.    

Figure~5 shows ionization histories, quantified by $Q_{\rm HII}(z)$ and $\tau_e(z)$, for various values of clumping 
factor, escape fraction, IGM temperature, and LyC-production efficiencies ($Q_{\rm LyC}$).   We adopt the SFR history
from Trenti \etal\ (2010) with an evolving luminosity function.   Our standard model adopts $C_H = 3$, $f_{\rm esc}=0.2$, 
$T_4 = 2$,  and $Q_{\rm LyC} = 0.004$ ($4\times10^{60}$ photons/M$_{\odot}$).   Different curves show the effects of 
changing $C_H$ and $f_{\rm esc}$, including two models in which these parameters evolve with redshift.    With these 
parameters, we are typically able to complete reionization by $z \approx 7$, consistent with observations of 
Gunn-Peterson troughs, redshift evolution in \Lya\ emitters, and IGM neutral fraction.    
The model with a constant $f_{\rm esc} = 0.05$ does not complete the ionization until $z \approx 4$, which is
far too late.  Thus, we be believe that LyC escape fractions must be considerably larger ($\geq20$\%) perhaps evolving 
to higher values at $z > 6$ shown by green and magenta curves.  

Figure 6 compares the dependence of $Q_{\rm HII}(z)$ and $\tau_e(z)$ on SFR histories, computed with an evolving
luminosity function (Trenti \etal\ 2010) and new calculations (Haardt \& Madau 2012) of the star formation rate density, 
$\dot{\rho}_{\rm SFR}(z)$.   Figure 7 compares the effects of two choices of model atmospheres, with a fixed SFR history 
from Trenti \etal\  (2010).  The on-line calculator provides ionization fractions, $Q_{\rm HII}(z)$ and $Q_{\rm HeIII}(z)$, 
together with CMB optical depth, $\tau_e(z)$.   In the simulator, users can select other IGM parameters and SFR histories.  
The main difference between the two SFR models lies in the assumptions at $z>8$. Haardt \& Madau (2012) rely on an 
empirical extrapolation of the star formation rate as a function of redshift, while Trenti \etal\  (2010) adopt a physically 
motivated model based on the evolution of the dark-matter halo mass function. The two approaches are similar at 
$z \lsim 8$, but differ significantly at higher redshift, where an empirical extrapolation does not capture the sharp drop 
in the number density of galaxies observed at $z \sim 10$ (see Figure 8 in Oesch \etal\ 2011). As a consequence, the 
reionization history from the Haardt \& Madau (2012) model is more extended at high $z$, especially when the 
efficiency of reionization is increased because of evolving clumping factor and escape fraction (our preferred models,
green and magenta curves in Fig.\ 5). The two models yield quite different predictions for the duration of reionization,   
defined as the redshift interval, $\Delta z$,  over which $Q_{\rm HII}$ evolves from 20\% to 80\% ionized.   Our preferred 
SFR models (green and magenta lines in Fig.\ 5) have $\Delta z \approx 3$ (from $z \approx 10.5$ to $z \approx 7.5$),
 whereas the Haardt-Madau SFR histories exhibit a more extended interval, $\Delta z \approx 6$ (from $z \approx 13$ to 
 $z \approx 7$).   This difference in $\Delta z$ could be tested by upcoming 21-cm experiments; see Bowman \& Rodgers 
 (2010) for an initial constraint, $\Delta z > 0.06$.  
 
Figure 8 illustrates a third constraint on the reionization epoch (Pritchard \etal\ 2010;  Lidz \etal\  2011) comparing SFR 
histories with estimates of the ionizing background at $z = 5.5 \pm 0.5$ (Fan \etal\ 2006;  Bolton \& Haehnelt  2007;  
Haardt \& Madau 2012).  The LyC co-moving emissivity (in photons s$^{-1}$~Mpc$^{-3}$), defined as
$\dot{n}_{\rm LyC} = \dot{\rho}_{\rm SFR}  f_{\rm esc} Q_{\rm LyC}$,  can be related to the ionizing background at 
$z = 5-6$, using recent estimates of the hydrogen photoionization rate, 
$\Gamma_{\rm HI}(z)$, from Haardt \& Madau (2012) and the LyC mean free path, $\lambda_{\rm HI}$, from Songaila \& 
Cowie (2010).   The LyC emissivity is proportional to the star formation rate density, computed from our halo mass function 
model (Trenti \etal\ 2010), integrated down to absolute magnitudes $M_{\rm AB} = -18$ (Bouwens \etal\  2011a) or to 
$M_{\rm AB} = -10$, the faint limit suggested by Trenti \etal\ (2010).   We adopt a fiducial LyC production 
parameter $Q_{\rm LyC} = 0.004$, corresponding to $4\times10^{60}$ LyC photons produced per $M_{\odot}$ of star 
formation, and we use two different models for LyC escape fraction, $f_{\rm esc}$ (constant at 20\% and varying with 
redshift).  For quantitative values, we assume an ionizing background with specific intensity, 
$J_{\nu} = J_0 (\nu / \nu_0)^{-\alpha}$  (in units erg~cm$^{-2}$~s$^{-1}$~sr$^{-1}$~Hz$^{-1}$) with power-law index 
$\alpha \approx 2$ at energies above $h \nu_0 = 1$ ryd.   The hydrogen photoionization rate is 
$\Gamma_{\rm HI} =  [4 \pi J_0 \sigma_0/ h (\alpha + 3)]$ for a hydrogen photoionization cross section 
$\sigma_{\nu} \approx \sigma_0 (\nu / \nu_0)^{-3}$ with $\sigma_0 = 6.3\times10^{-18}$~cm$^2$.  For 
this spectrum, the frequency-integrated ionizing intensity is $J_{\rm tot} = J_0 \nu_0 / (\alpha-1)$, and the LyC photon 
flux (photons~cm$^{-2}$~s$^{-1}$) integrated over all solid angles is $\Phi_{\rm LyC} = (4 \pi J_0/h \alpha)$.  Because 
we analyze the {\it photon} mean-free path, we normalize $J_0$ to $\Phi_{\rm LyC}$ and $\Gamma_{\rm HI}$,
\begin{equation}
   J_0 = \frac {h \alpha \Phi_{\rm LyC}} {4 \pi } = \frac {h (\alpha + 3)} {4 \pi \sigma_0}  \; \Gamma_{\rm HI}  \; . 
\end{equation}
By approximating $\Phi_{\rm LyC}$ as the product of the LyC emissivity and mean-free path, we arrive at the calibrations:
\begin{eqnarray}
   J_{\rm tot} &=&  \frac {h \nu_0} {4 \pi \sigma_0}  \frac {(\alpha+3)} {(\alpha - 1)}  \;  \Gamma_{\rm HI}
      = (6.88\times10^{-7}~{\rm erg~cm}^{-2}~{\rm s}^{-1}~{\rm sr}^{-1}) \left[ \frac {\Gamma_{\rm HI}} 
         {5 \times 10^{-13}~{\rm s}^{-1}} \right]     \\
   \Phi_{\rm LyC} &=&  \frac {(\alpha+3)}{\alpha}  \frac {\Gamma_{\rm HI}} { \sigma_0}  
        = (1.98 \times 10^5~{\rm photons~cm}^{-2}~{\rm s}^{-1}) 
           \left[ \frac {\Gamma_{\rm HI}} {5 \times 10^{-13}~{\rm s}^{-1}} \right]   \;  .  
\end{eqnarray}   
Finally, we relate $\Phi_{\rm LyC}$ to the star-formation rate $\dot{\rho}_{\rm SFR}$ and ionization rate
$\Gamma_{\rm HI}$, 
\begin{equation}
    \dot{\rho}_{\rm SFR} =  (0.056~M_{\odot}~{\rm yr}^{-1}~{\rm Mpc}^{-3}) 
             \left[ \frac {\Gamma_{\rm HI}}{5 \times 10^{-13}~{\rm s}^{-1}} \right] \left[ \frac {0.1}  {f_{\rm esc}} \right] 
             \left[ \frac  {0.004}{Q_{\rm LyC}} \right] \left[ \frac {9.8~{\rm pMpc}} {\lambda_{\rm HI}} \right] 
             \left[ \frac {6.5}{1+z} \right]^{3}       \; . 
\end{equation}
Here, we have scaled the parameters to the same values assumed by Lidz \etal\ (2011), namely 
$\Gamma_{\rm HI} = 5 \times 10^{-13}~{\rm s}^{-1}$, $f_{\rm esc} = 0.1$,  an ionizing spectral index $\alpha = 2$,
and redshift $z = 5.5$, at which Songaila \& Cowie (2010) fit  $\lambda_{\rm HI} \approx 9.8$ proper Mpc.   
Our parameter $Q_{\rm LyC} = 0.004$ corresponds to the Lidz \etal\ (2011) LyC photon production calibration,
$10^{53.1}~{\rm s}^{-1}$ per $M_{\odot}~{\rm yr}^{-1}$ of star formation.  However, our coefficient, 
$0.056~M_{\odot}~{\rm yr}^{-1}~{\rm Mpc}^{-3}$, is slightly larger than their value,
$0.039~M_{\odot}~{\rm yr}^{-1}~{\rm Mpc}^{-3}$, an effect that may arise from our different method of 
relating $\Gamma_{\rm HI}$ to the ionizing radiation field and SFR.  

More careful examination of these parameters suggests that, at $z \approx 5.5$, the hydrogen ionization 
rate  $\Gamma_{\rm HI} \approx 3.6 \times 10^{-13}~{\rm s}^{-1}$, given in Table 3 of Haardt \& Madau (2012), 
and the observed mean free path, $\lambda_{\rm HI}$, at $z = 5.5$ may be closer to 6 proper Mpc (see Figure 10 
of Songaila \& Cowie 2010).   Rescaling to those two parameters, we find a similar coefficient of 
$0.066~M_{\odot}~{\rm yr}^{-1}~{\rm Mpc}^{-3}$.   This SFR density at $z = 5.5 \pm 0.5$ (see also Figure 8) is 
comparable to the critical value needed to maintain reionization at $z = 7$, but the observed rates and mean free 
paths are declining rapidly with redshift.  All three constraints on SFR suggest that full reionization is more likely 
to occur at redshift $z_{\rm rei} \approx 7$ than at $z_{\rm rei} = 10$.

 
\begin{figure}[h]
\begin{center}
\includegraphics[angle=270,scale=0.55]{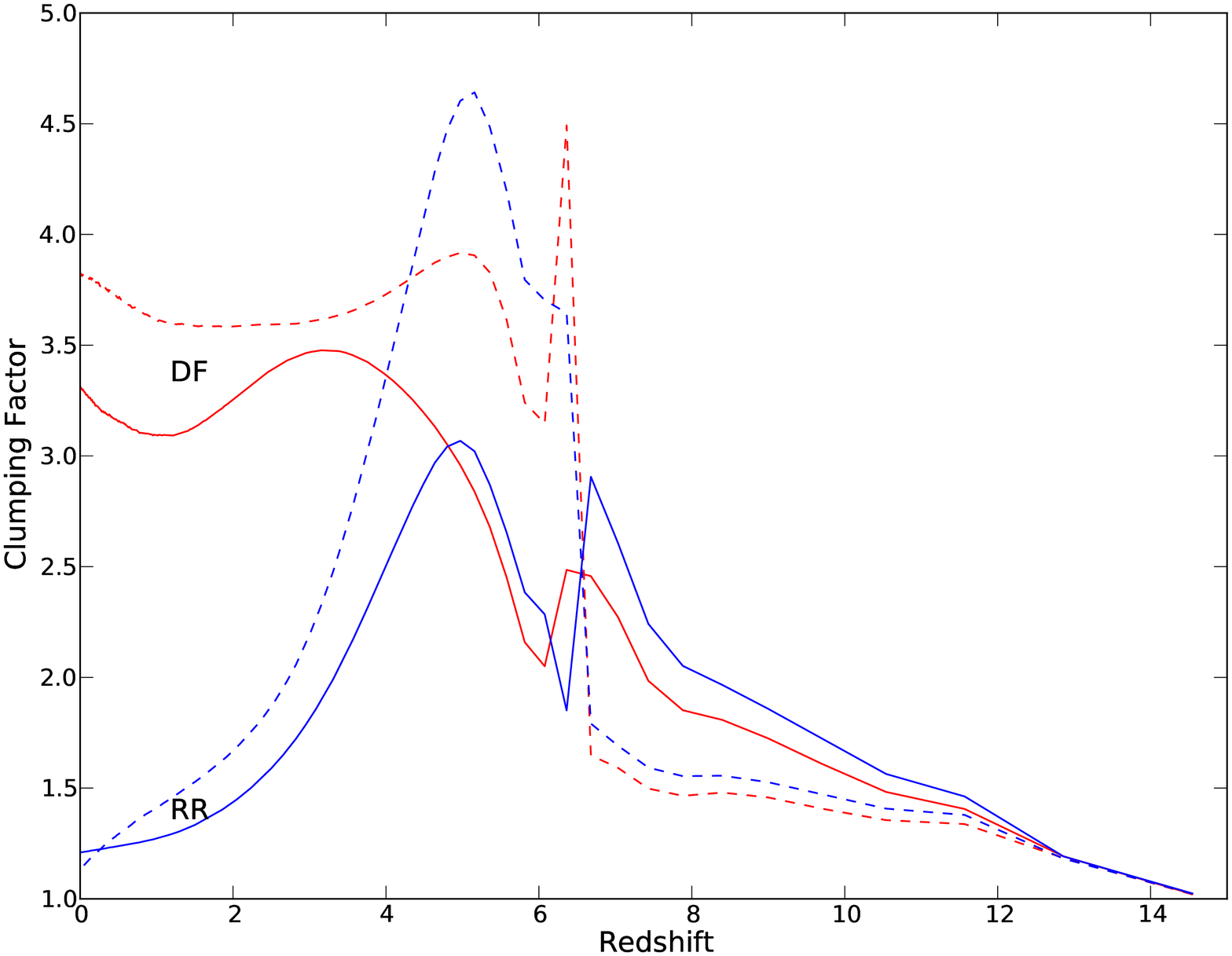}
\caption{Comparison of two methods for calculating the clumping factor for simulation $50\_1024\_7$ 
(UV background turned on at $z=7$) with different weights.  Red lines:  clumping calculated via density field 
(DF) method.  Blue lines:  clumping via recombination rate (RR) method (see Section 2.2).  Solid lines 
correspond to weighting by volume, and dashed lines correspond to weighting by baryon overdensity.  
We believe the RR method, with volume weighting, is a more accurate measure of clumping. }
\label{fig:RRCF_vs_clump_both}
\end{center}
\end{figure}

 
\begin{figure}[h]
\begin{center}
\includegraphics[angle=270,scale=0.55]{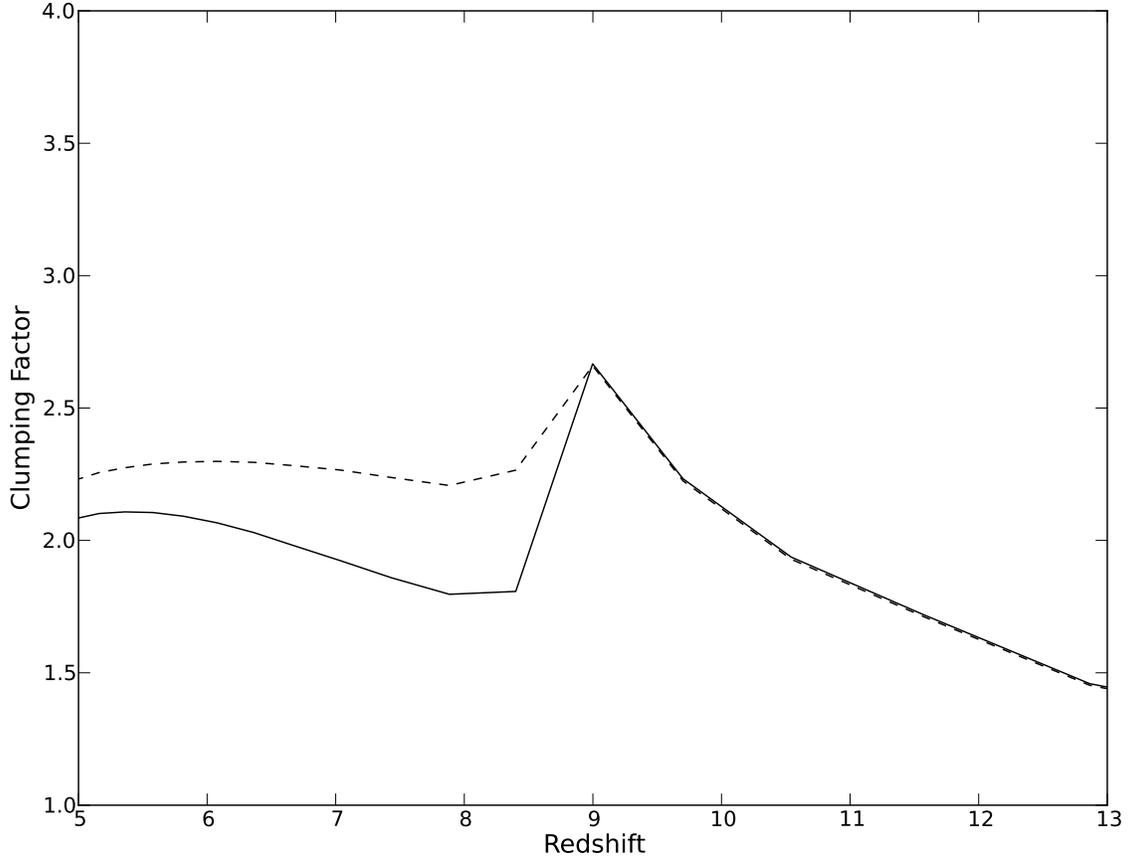}
\caption{Effects of different ``density cuts" on the clumping factor from our $1536^3$ simulation,
using the RR method in Equation (15).  
Solid line: clumping factor summed over cells with overdensity between $1 < \Delta_b <100$,
excluding low-density voids with $\Delta_b < 1$.  Dashed line:  including all cells with $\Delta_b < 100$.  
 }
\label{fig:RRCF_vs_clump_vol}
\end{center}
\end{figure}


 
\begin{figure}[h]
\begin{center}
\includegraphics[angle=270,scale=0.4]{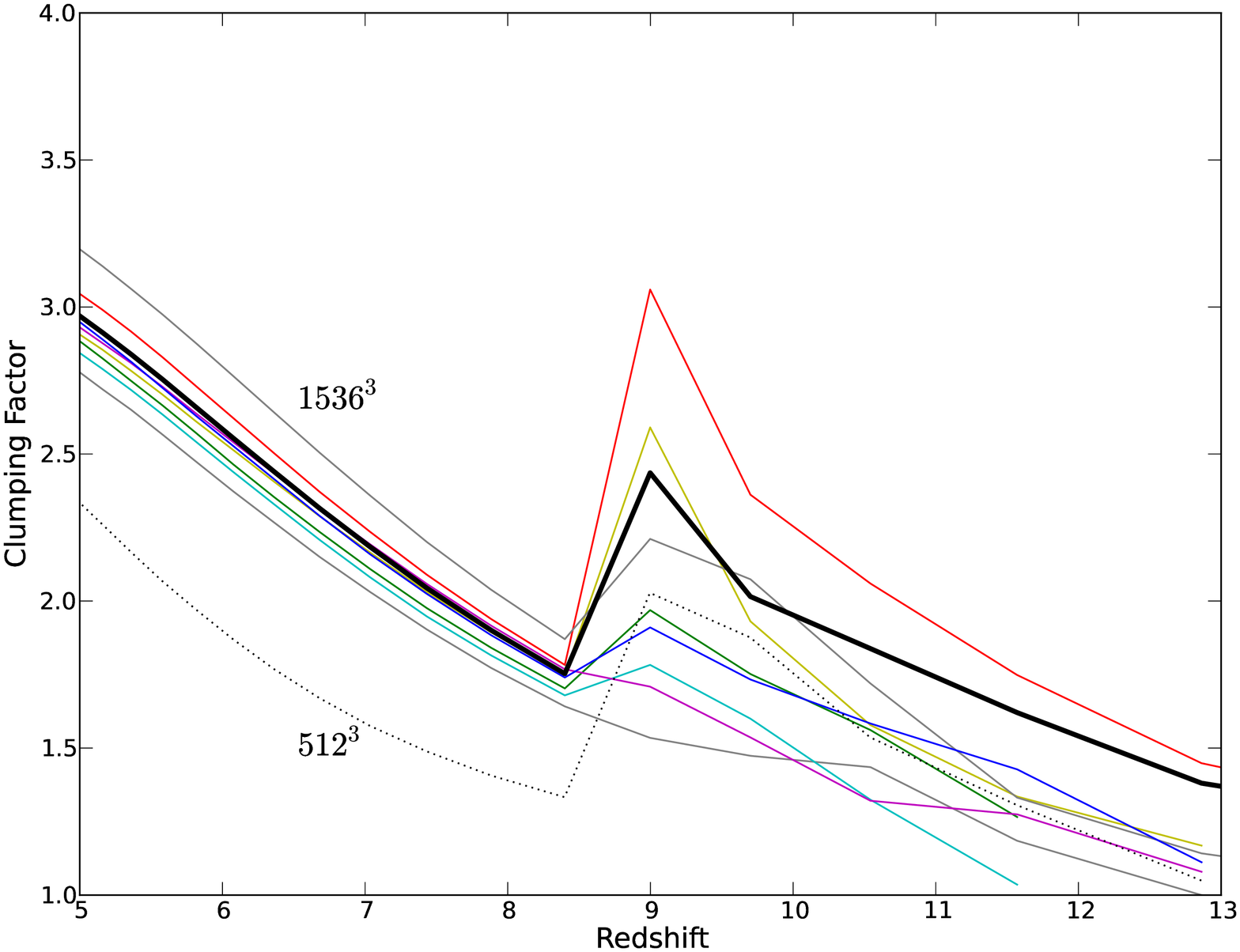}
\includegraphics[angle=270,scale=0.4]{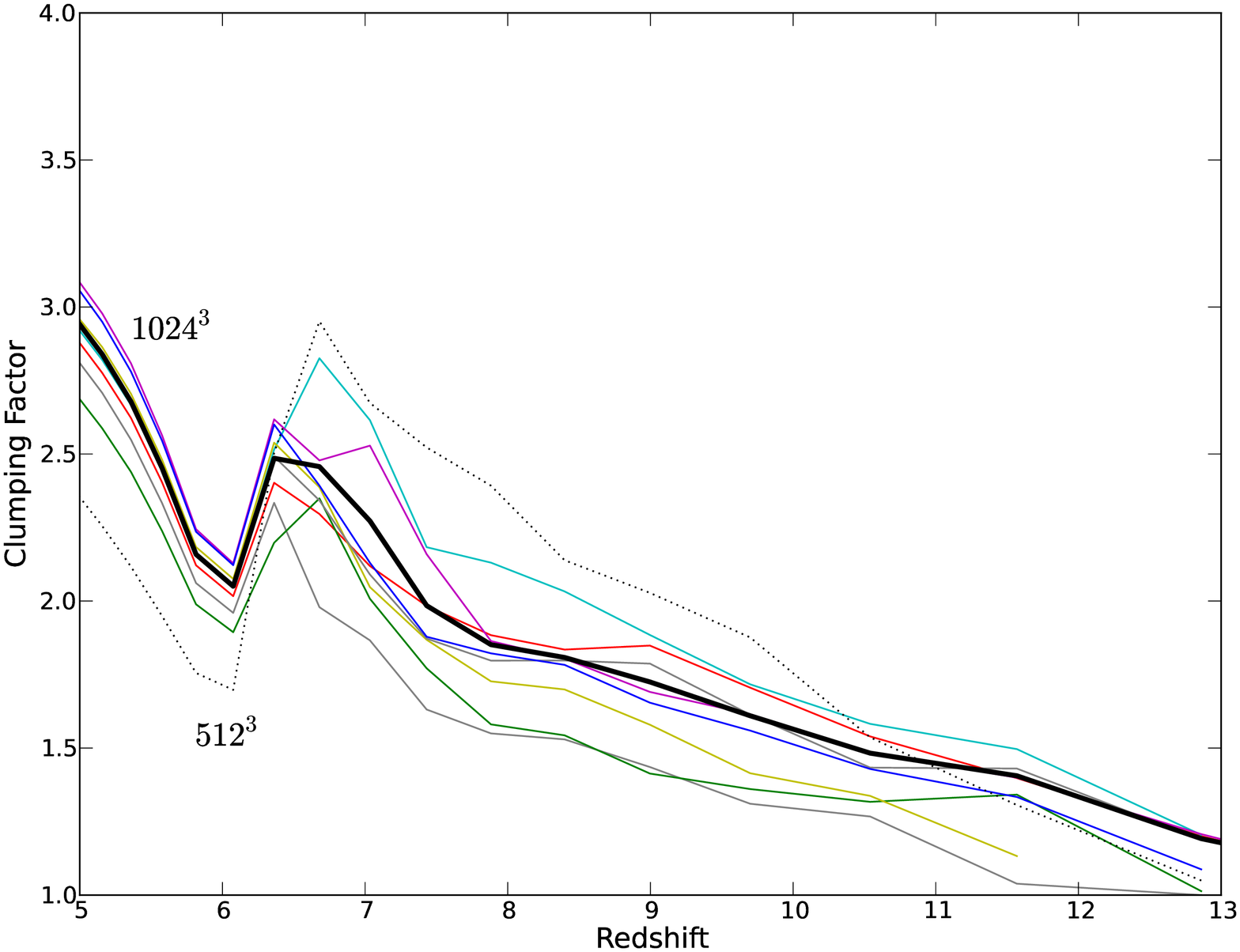}
\caption{Comparison of convergence and cosmic variance among simulations.  Values of $C_H$ agree
for $1024^3$ and $1536^3$ simulations, with variance computed over eight sub-volumes in our simulations.  
Top panel:  eight $768^3$ sub-volumes in $1536^3$ run.   Bottom panel:  eight $512^3$ sub-volumes in 
$1024^3$ run.   In each panel, our original $512^3$ run is shown as dotted black line, and values for
$1536^3$ and $1024^3$ runs as heavy black lines.   Large variations in $C_H$ at high redshifts arise from 
small numbers of cells prior to turn-on of the ionizing background at $z = 9$ (top) and $z = 7$ (bottom).  
For the $1536^3$ simulation (top panel), the clumping factor is well-fitted by $C_H(z) = (2.9) [(1+z)/6]^{-1.1}$ 
between $5 < z < 9$.  
 }
\label{fig:RRCF_vs_clump_vol}
\end{center}
\end{figure}



\begin{figure}[h]
\begin{center}
\includegraphics[angle=270,scale=0.55]{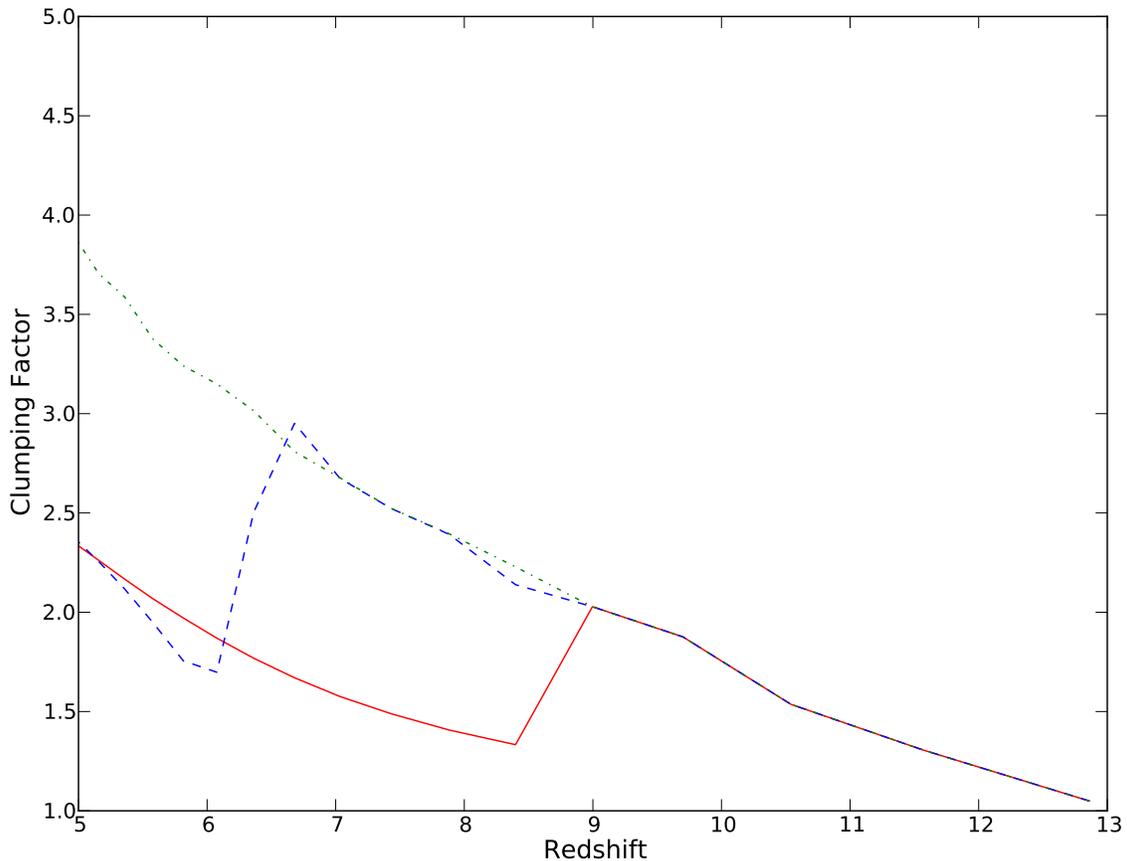}
\caption{Clumping factor calculated via the density field (DF) method computed from three $512^3$ IGM 
simulations, with sums weighted by volume (see eq. [14]).   Red solid line has UV background 
(Haardt \& Madau 2001) turned on at $z=9$ (simulation $50\_512\_9$), blue dashed line at $z=7$
(simulation $50\_512\_7$),  and green dot-dashed line has no background (simulation $50\_512\_0$).
We observe two tracks for $C_H(z)$:  a high-track for redshifts above turn-on of the UV
background, and a lower track after turn-on.  During most of the reionization epoch, from $6 < z < 12$,
$C_H$ lies between 1.5 and 3.  
}
\label{fig:RRCF_vol_background}
\end{center}
\end{figure}



\begin{figure}[h]
\begin{center}
\includegraphics[angle=270,scale=0.55] {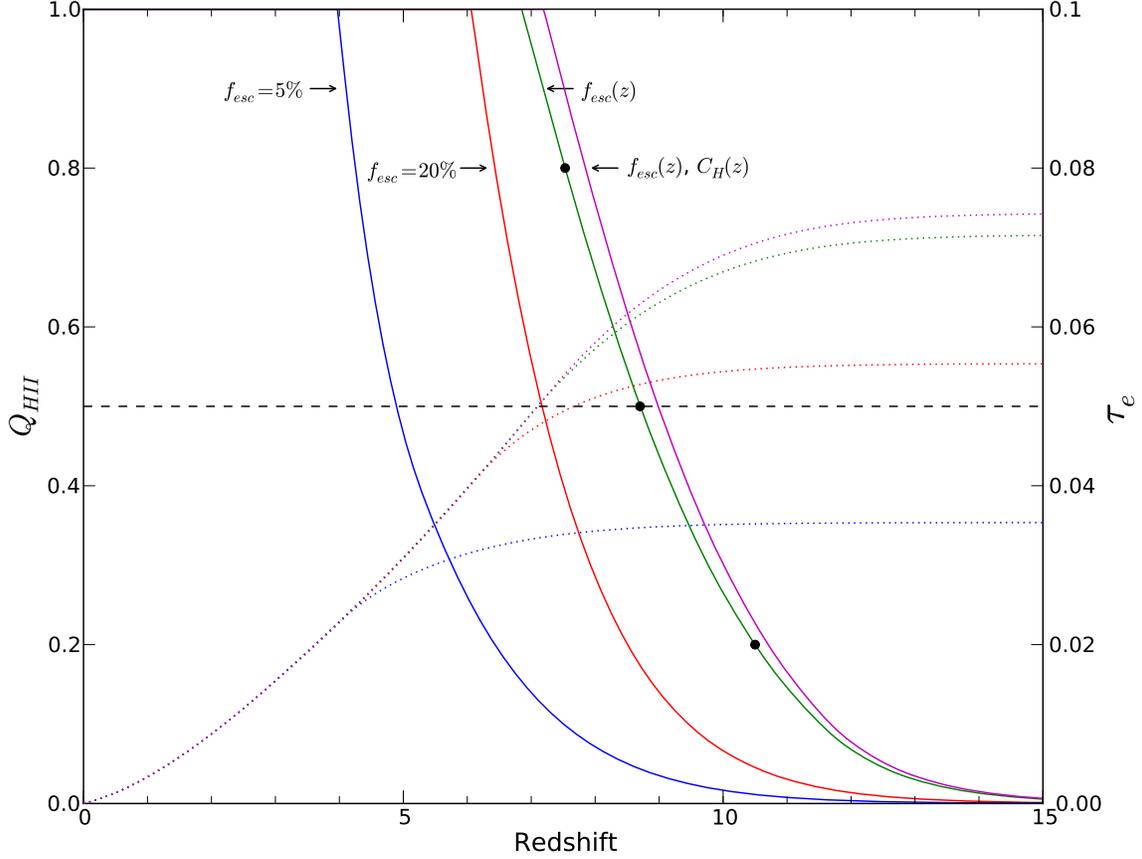} 
\caption{History of ionization fraction $Q_{\rm HII}(z)$ and CMB optical depth $\tau_e(z)$  versus redshift 
(eqs.\  [3] and [17]) computed with $T_e = 2 \times 10^4$ K, $Q_{\rm LyC} = 0.004$
($4 \times 10^{60}$ photons per $M_{\odot}$) and using SFR history and evolving luminosity function from 
Trenti \etal\ (2010) integrated down to $M_{\rm AB} = -10$.  
The IGM is assumed to be fully ionized when $Q_{\rm HII} = 1$.   
Blue line: constant $C_H = 3$ and $f_{\rm esc} = 0.05$ (this model reionizes too late to be viable).  
Red line: constant $C_H = 3$ and $f_{\rm esc} = 0.2$.   
Green line:  constant $C_H = 3$ but redshift-dependent $f_{\rm esc} = 1.8 \times 10^{-4} (1 +z)^{3.4}$
(Haardt \& Madau 2012).    Magenta line:  redshift-dependent $C_H = 1+43z^{-1.71}$ (Pawlik \etal\ 2009) and 
$f_{\rm esc} =1.8\times10^{-4} (1 +z)^{3.4}$.  
Our two preferred models with variable $C_H$ or $f_{\rm esc}$ (green, magenta) naturally produce 
$z_{\rm rei} \approx 7$. Solid black circles indicate redshifts of 20\%, 50\%, and 80\% ionization;  the duration 
$\Delta z \approx 3$ is defined between 20\% and 80\% points.  
 }
\label{fig:ionization_history}
\end{center}
\end{figure}



\begin{figure}[h]
\begin{center}
\includegraphics[angle=270,scale=0.35]{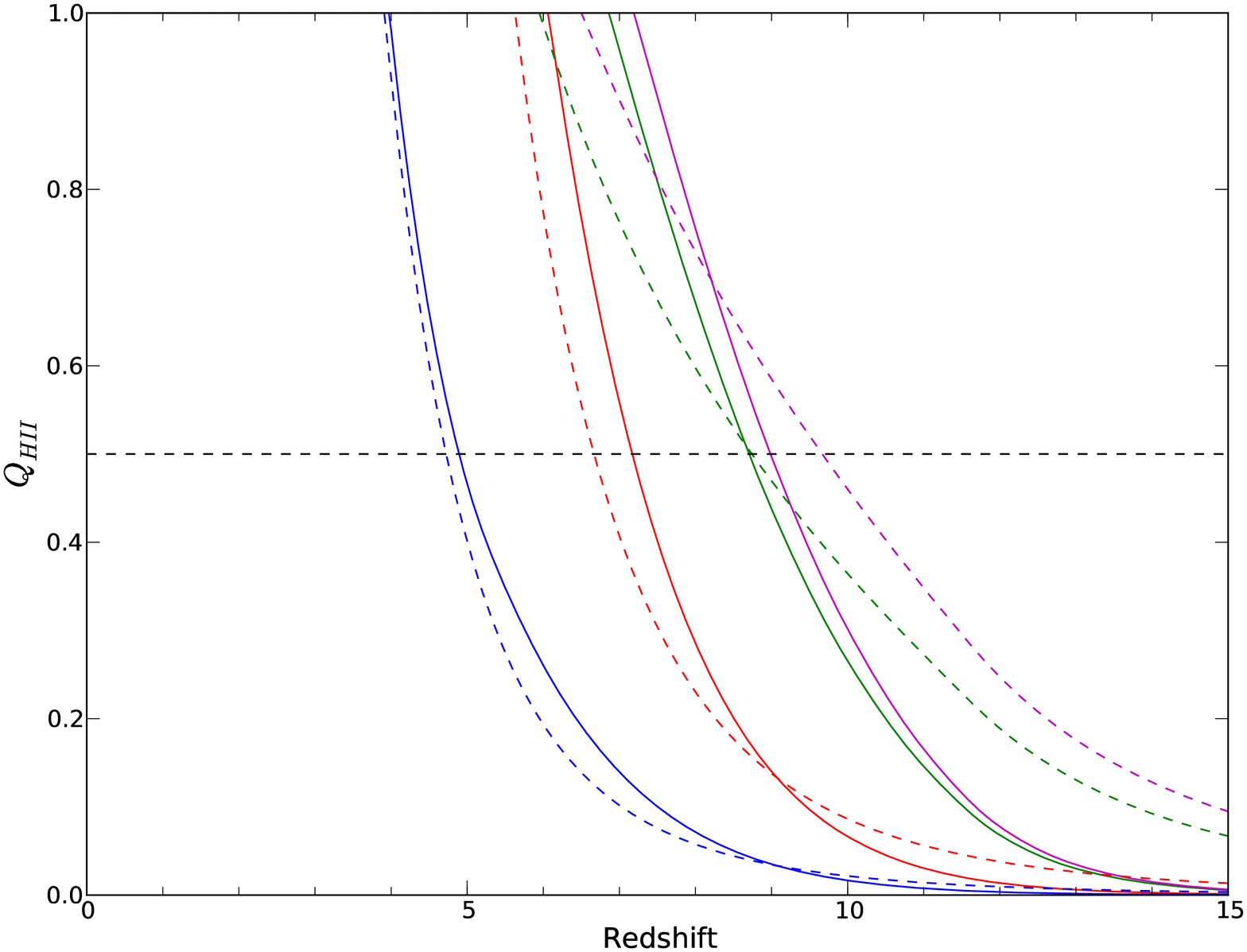} 
\includegraphics[angle=270,scale=0.35]{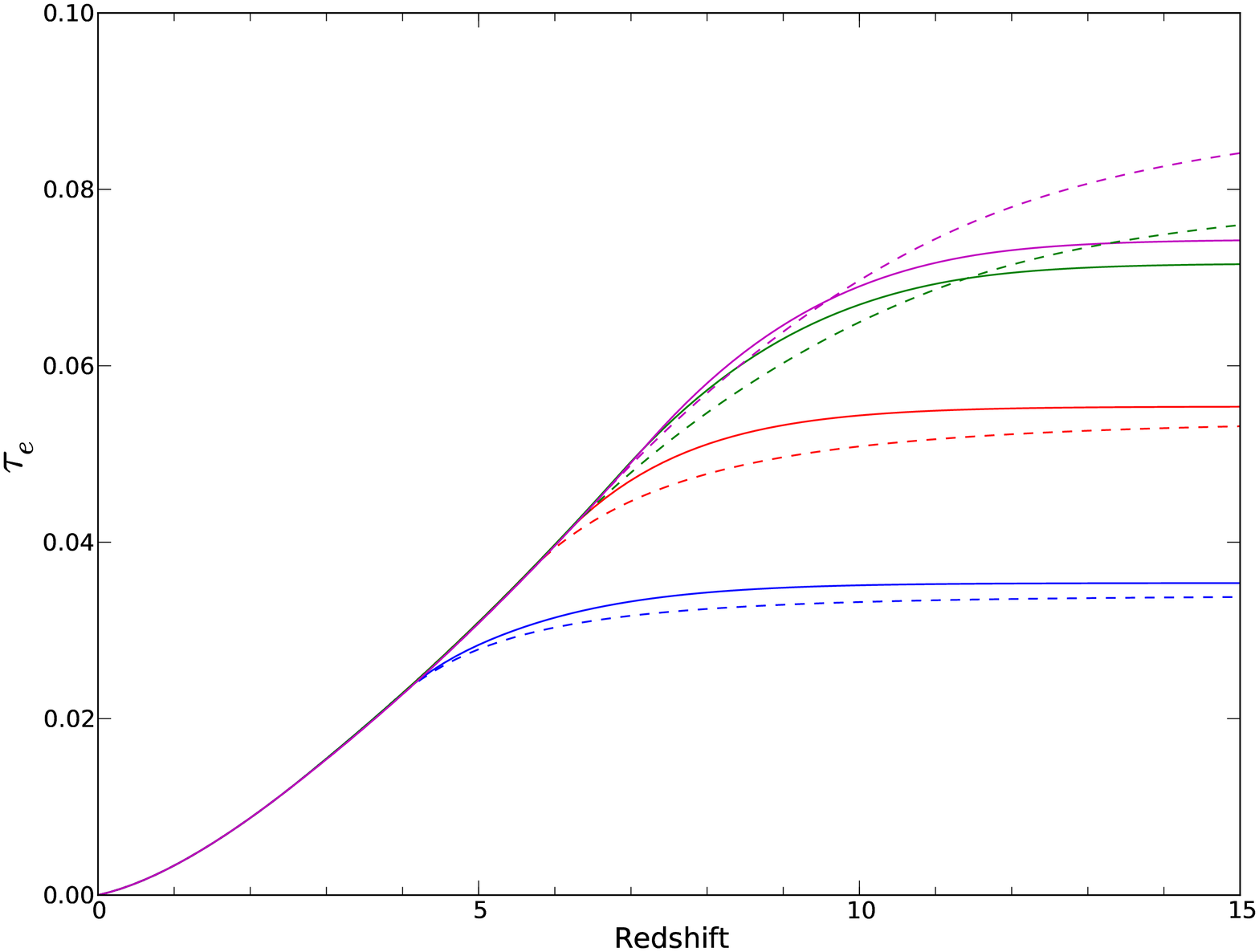} 
\caption{ Top: Volume filling factors, $Q_{\rm HII}$, with same color codes as Fig.\ 5, comparing two models for 
SFR history.   Solid lines show SFR history from Trenti \etal\ (2010), and dashed lines show SFR from equation (53)
in Haardt \& Madau (2012), which is larger and more extended at higher redshifts, particularly for $z > 8$.  
Bottom:  Corresponding CMB optical depths $\tau_e(z)$.   Additional optical depth may arise from sources at $z > 7$.  
}
\end{center}
\end{figure}



\begin{figure}[h]
\begin{center}
\includegraphics[angle=270,scale=0.55]{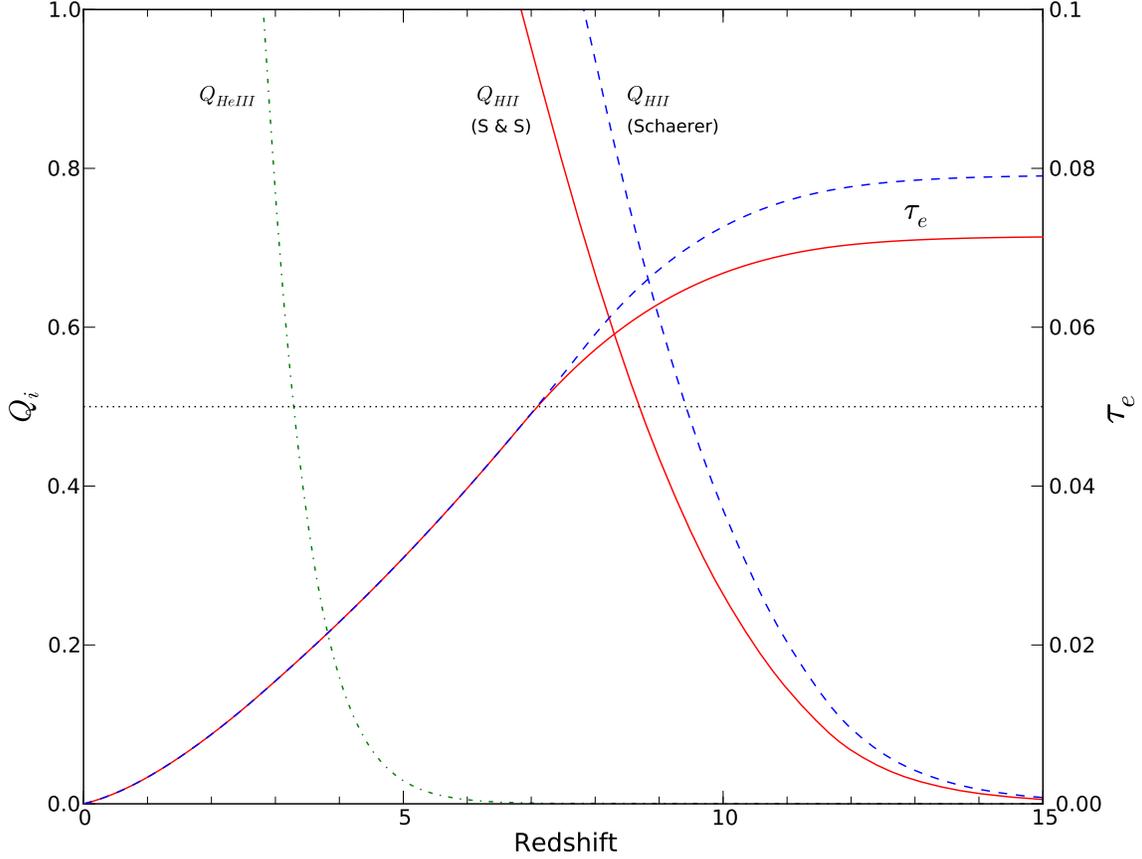} 
\caption{Example of the output for $Q_{\rm HII}(z)$, $Q_{\rm HeIII}(z)$, and $\tau_e(z)$ from our
on-line SFR/reionization simulator.  Solid lines show model atmospheres labeled as Sutherland \& Shull 
(S\&S), while dashed lines are for Schaerer (2002, 2003), assuming $Z = 0.02 Z_{\odot}$.   
Models assume SFR history 
of Trenti \etal\ (2010) and parameters with constant $C_H = 3$ and redshift-dependent 
$f_{\rm esc} = 1.8 \times 10^{-4} (1 +z)^{3.4}$ (Haardt \& Madau 2012).    
Dot-dashed line shows \HeIII\ ionization history, $Q_{\rm HeIII}(z)$, computed for
QSO (4-ryd continuum) ionization as described in Section 3.3.  In these models, \HI\
is fully ionized by $z \approx 7$ and \HeIII\ by $z \approx 2.7$.  Additional optical depth
may arise from sources at $z > 7$.  
}
\end{center}
\end{figure}


\begin{figure}[h]
\begin{center}
\includegraphics[angle=270,scale=0.40]{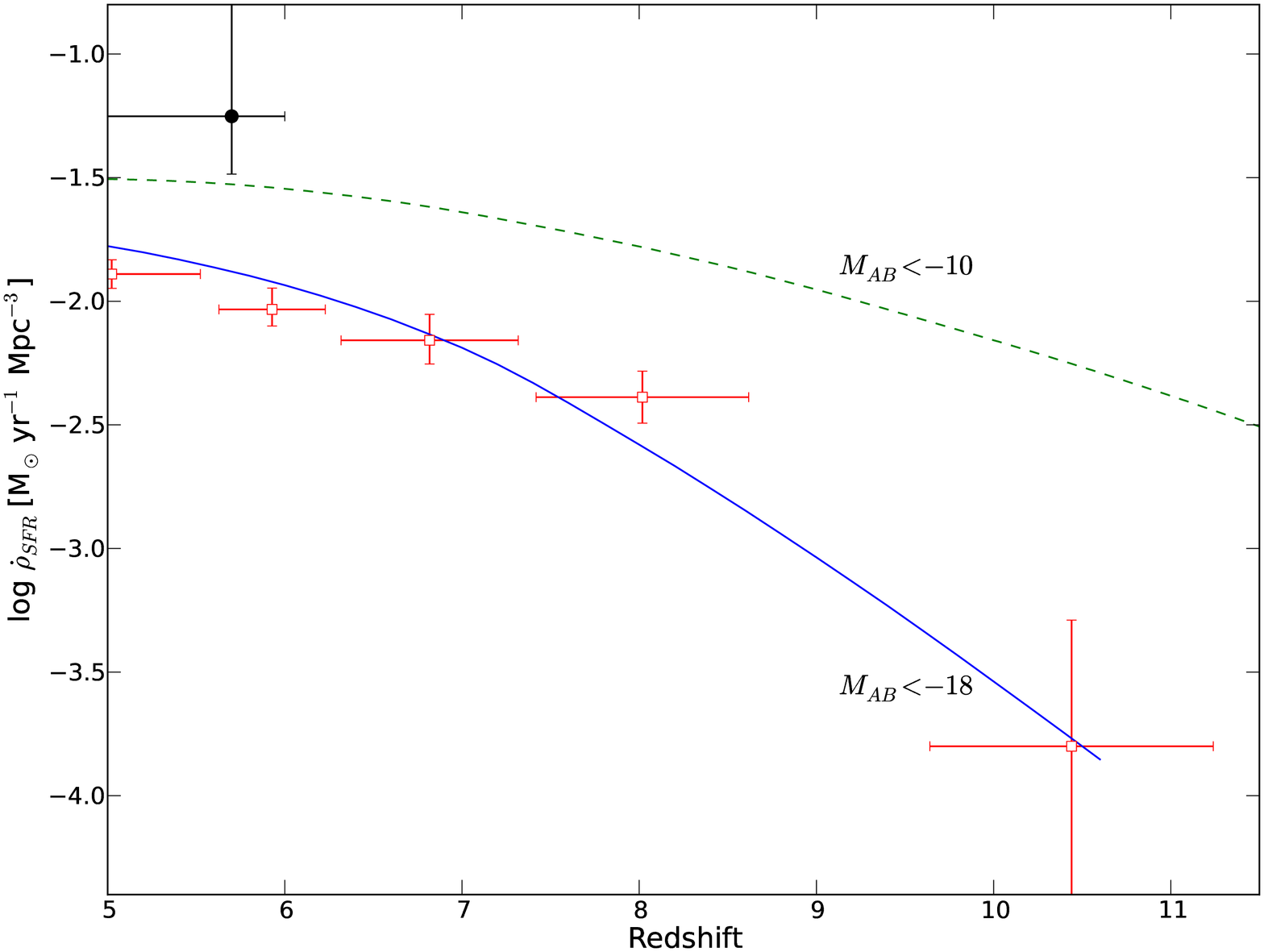} 
\includegraphics[angle=270,scale=0.40]{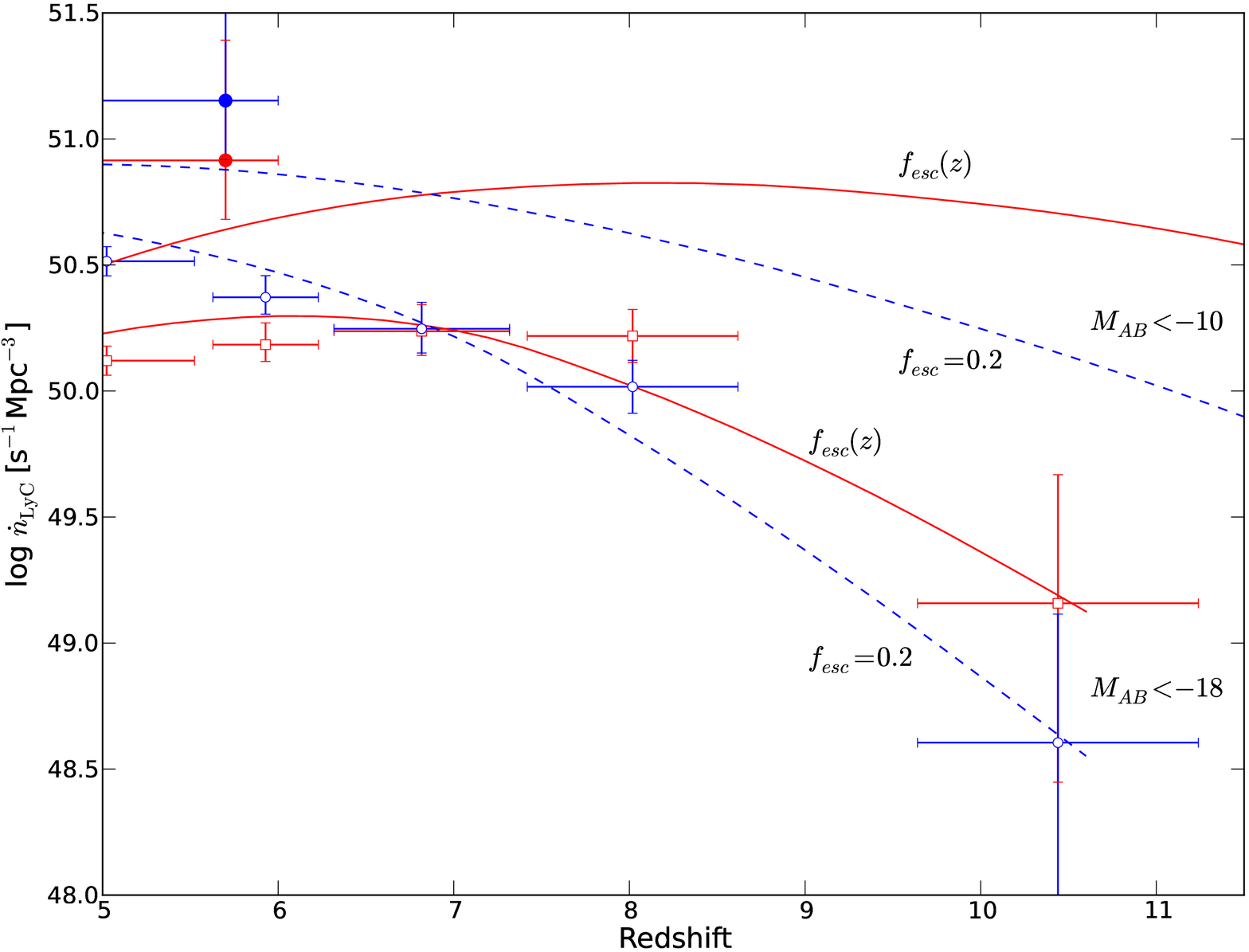} 
\caption{ Top:  Co-moving SFR density ($\dot{\rho}_{\rm SFR}$) from halo mass function model
(Trenti \etal\  2010) integrated for galaxies down to $M_{\rm AB} = -18$ (solid blue line) and 
to $-10$ (dotted green line).   Red data points are from Bouwens \etal\ (2011a) for galaxies down to 
$M_{\rm AB} = -18$.    Solid data point at $z \approx 5.5$ is from ionizing background constraint (Eq.\ 21),
using Haaradt \& Madau (2012) model of $\Gamma_H(z)$ and LyC mean free paths from
(Songaila \& Cowie 2010).  Bottom:  Co-moving emissivity, $\dot{n}_{\rm LyC}$, of LyC photons, 
corresponding to above SFRs, integrated to $M_{\rm AB} = -18$ and $-10$.  We use a LyC production
parameter $Q_{\rm LyC} = 0.004$ ($10^{53.1}$ LyC photons~s$^{-1}$ per $M_{\odot}$~yr$^{-1}$) 
and two different escape-fraction models:  $f_{\rm esc} = 0.2$ (blue curves and data points) and 
$f_{\rm esc}(z) = (1.8\times10^{-4})(1+z)^{3.4}$ (red curves and data points).  
}
\end{center}
\end{figure}

\section{DISCUSSION AND IMPLICATIONS}  

\noindent
The major results of our study can be summarized with the following points:
\begin{enumerate}

\item  We calculated the critical star formation rate required to maintain a photoionized IGM,
incorporating four free parameters ($C_H$, $f_{\rm esc}$, $T_e$, $Q_{\rm Lyc}$) that control
the rates of LyC photon production and radiative recombination.  Our best estimate at $z = 7$ is
$\dot{\rho}_{\rm crit} \approx (0.018 \, M_{\odot}~{\rm yr}^{-1}~{\rm Mpc}^{-3})[(1+z)/8]^3 
(C_H/3)(0.2/f_{\rm esc}) T_4^{-0.845}$  for fiducial values of IGM clumping factor $C_H  =  3$, 
LyC escape fraction $f_{\rm esc} = 0.2$, temperature $T_e = 10^4$~K, and standard IMFs 
and low-metallicity stellar atmosphere ($Q_{\rm LyC} = 0.004$).  

\item An epoch of full reionization at $z_{\rm rei} \approx 7$ is consistent with recent optical/IR 
measurements of SFR history and a rising IGM neutral fraction at $z = 6-8$, marking the tail end of 
reionization.   These observations include the decrease in numbers of
high-$z$ galaxies and \Lya\ emitters and IGM damping-wing intrusion into the \Lya\ transmission
profiles in high-$z$ QSOs.   

\item Our newly developed SFR and reionization calculator, now available on-line at\\
\url{http://casa.colorado.edu/~harnessa/SFRcalculator}, allows users to calculate the ionization history,
critical SFR, and CMB optical depth, and to assess whether observed SFRs, IMFs, and other parameters 
are consistent with IGM reionization.  

\item  Reconciling late reionization at $z_{\rm rei} \approx 7$ with $\tau_e = 0.088\pm0.015$ of the CMB
likely requires an epoch of partial ionization.   A fully ionized IGM back to $z_{\rm rei} = 7$ produces 
$\tau_e \approx 0.050$, and an additional optical depth, $\Delta \tau_e \approx 0.01-0.04$, may
arise from early sources of UV/X-ray photons at $z > 7$.  Alternatively, one can appeal to the likelihood
contours from WMAP, which allow optical depths as low as $0.06$ (95\% C.L.)  The {\it Planck} experiment 
may clarify the situation in several years.  

\item  If the EoR is more complex, as we suggest, then redshifted 21-cm experiments should focus on the 
interval $7.5 < z < 10.5$, corresponding to frequencies 124--167 MHz, for the maximum signal of IGM heating 
($T_b \approx 27$~mK) produced when the hydrogen neutral fraction 
$x_{\rm HI} \approx 0.5$ (Pritchard \etal\  2010).  

\end{enumerate}

\noindent
We conclude by speculating about which future observations can best constrain the EoR.   Additional data
from from the  {\it Planck} mission\footnote{See mission website {\tt http://www.rssd.esa.int/Planck}} 
should provide confirmation of the CMB optical depth with smaller error bars.  This information will constrain the 
additional amount of ionization at $z > 7$ and narrow the range $\Delta \tau_e = 0.01-0.04$ produced by 
high-redshift sources in the partially ionized IGM.   Ongoing surveys for high-$z$ galaxies, \Lya\ emitters, and 
QSO near-zone sizes will better quantify the rise of neutral fraction, $x_{\rm HI}$ at $z > 6.5$.    On the theoretical 
front, we are running larger IGM simulations on $1536^3$ and $2048^3$ unigrids and will add discrete sources 
of ionizing radiation and radiative transfer in order to capture the heating and clumping more accurately.   We will 
also include source turn-on at $z > 9$ to test the decrease and recovery of $C_H$, as seen in Figures 3 and 4.   
Finally, we plan to carry out more detailed modeling of the \HI\ (21-cm) signal, coupled to the kinetic and spin 
temperatures driven by heating at $z > 7$.  As described earlier, the duration of the reionization transition and the 
21-cm emission during the heating phase could provide discriminants of various SFR histories at $z > 6$.


\acknowledgments

This work was supported by grants to the Astrophysical Theory Program (NNX07-AG77G  from NASA and 
AST07-07474 from NSF) at the University of Colorado Boulder.  We thank Joanna Dunkley, Chris Carilli, 
Richard Ellis, and Piero Madau for useful discussions on reionization and CMB optical depth statistics.  We
are grateful to the referee for suggesting additional model comparisons with the ionizing 
background at $z = 5-6$.  




\begin{deluxetable}{cccccccc}
\tabletypesize{\scriptsize}
\tablecaption{Sample values of $Q_{\rm LyC}$ and $\dot{\rho}_{\rm crit}$ for different IMFs}
\tablewidth{0pt}
\tablehead{
\colhead{$M_{\rm min} \left(M_\odot\right)$} & \colhead{$M_{\rm max} \left(M_\odot\right)$} & \colhead{$\alpha$}
& \colhead{$Z$} & \colhead{$Q_{\rm LyC}$\tablenotemark{a}}& \colhead{$Q_{\rm LyC}$\tablenotemark{b}}
& \colhead{$\dot{\rho}_{\rm crit}$\tablenotemark{a}}& \colhead{$\dot{\rho}_{\rm crit}$\tablenotemark{b}}}
\startdata
0.1    &    100    &    2.35          &    $Z_\odot$               &    0.00236     &    0.00286    &    0.0306    &    0.0253      \\
0.1    &    100    &    2.35          &    $0.02Z_\odot$       &    0.00397     &    0.00558    &    0.0181    &    0.0129      \\
0.1    &    100    &    2.35          &    $0$                           &    0.00401     &    0.00752    &    0.0180    &    0.0096      \\
0.1    &    100    &    2.35          &    $0.2 Z_\odot$         &    0.00365     &   $\cdots$    &    0.0197    &    $\cdots$    \\
0.1    &    100    &    2.35          &    $0.1 Z_\odot$         &    0.00383     &    $\cdots$   &    0.0188    &    $\cdots$    \\
1.0    &    100    &    2.35          &    $0.1 Z_\odot$         &    0.00976     &    $\cdots$   &    0.0074    &    $\cdots$    \\
0.1    &    100    &    2.00          &    $0.1 Z_\odot$         &    0.01267     &    $\cdots$   &    0.0057    &    $\cdots$     \\
0.1    &    200    &    2.35          &    $0.1 Z_\odot$         &    0.00543     &    $\cdots$    &    0.0133    &   $\cdots$      \\
\enddata
\tablenotetext{a}{Sutherland \& Shull unpublished model atmospheres }
\tablenotetext{b}{Schaerer model atmospheres}
\tablecomments{Production efficiency of ionizing (LyC) radiation, $Q_{\rm LyC}$, in units of $10^{63}$
photons/M$_{\odot}$.  Critical SFR, $\dot{\rho}_{\rm crit}$, is given in units of M$_{\odot}$ yr$^{-1}$ Mpc$^{-3}$ 
(comoving) assuming $C_H = 3$, $f_{\rm esc} = 0.2$, and $T_4 = 1$.
}
\end{deluxetable}



\begin{deluxetable}{lcccc}
\tabletypesize{\scriptsize}
\tablecaption{Parameters of Simulations} \label{tbl:sims}
\tablewidth{0pt}
\tablehead{
\colhead{Run} & \colhead{$l$} & \colhead{$N_{\text{\rm cells}}^{1/3}$} & \colhead{$z_{\text{UV}}$} & \colhead{Relative Strength} \\ 
\colhead{} & \colhead{{($h^{-1}$ Mpc)}} & \colhead{} & \colhead{} & \colhead{{of Background}}}
\startdata
$50\_1024\_7$\tablenotemark{a}			        &	50	&	1024 	&	7	&	1\\
$50\_512\_7$\tablenotemark{b}			         &	50	&	512		&	7	&	1\\
$50\_512\_9$								&	50	&	512		&	9	&	1\\
$50\_512\_9\_2$							&	50	&	512		&	9	&	2\\
$50\_512\_0$								&	50	&	512		&	N/A	&	0\\
$50\_1536\_9$								&	50	&    1536 		&	9	&	1\\
\enddata
\tablenotetext{a,b}{Simulations~$50\_1024\_2$, $50\_512\_2$ from Smith \etal\  (2011).} 
\tablecomments{$z_{\text{UV}}$ is the redshift at which the ionizing background radiation 
(Haardt \& Madau 2001) is turned on, ramping up to a constant value by $z = z_{\text{UV}} - 1$}
\end{deluxetable}


\end{document}